\documentclass{article}

\usepackage{arxiv}

\usepackage[utf8]{inputenc} 
\usepackage[T1]{fontenc}    
\usepackage{hyperref}       
\usepackage{url}            
\usepackage{booktabs}       
\usepackage{nicefrac}       
\usepackage{microtype}      
\usepackage{lipsum}		
\usepackage{graphicx}
\usepackage{natbib}
\usepackage{doi}

\usepackage{graphicx}
\usepackage{newtxtext}
\usepackage{newtxmath}
\usepackage{natbib}

\newcommand{\RomanNumeralCaps}[1]
\linenumbers

\usepackage{graphicx,color,epsfig,soul,rotating,overpic,varwidth,xcolor}

\usepackage[export]{adjustbox}
\usepackage{subfigure}
\usepackage{subfigmat}
\usepackage{algorithm}
\usepackage[noend]{algpseudocode}
\usepackage{bm}
\usepackage{xfrac}
\usepackage{multicol}
\usepackage{multirow}
\usepackage{pbox}
\usepackage{tabularx}
\usepackage{caption}

\usepackage{booktabs}
\usepackage{tikz}
\usetikzlibrary{tikzmark}
\usetikzlibrary{calc}
\usetikzlibrary{arrows.meta}

\usepackage{lipsum}

\definecolor{blue2}{rgb}{0, 0.4470, 0.7410}
\definecolor{red2}{rgb}{0.8500, 0.1250, 0.0480} 
\definecolor{orange2}{rgb}{0.8500, 0.3250, 0.0980} 
\definecolor{yellow2}{rgb}{0.9290, 0.6940, 0.1250}
\definecolor{purple2}{rgb}{0.4940, 0.1840, 0.5560}
\definecolor{green2}{rgb}{0.4660, 0.6740, 0.1880}
\definecolor{ltblue2}{rgb}{0.3010, 0.7450, 0.9330}
\definecolor{dkred2}{rgb}{0.6350, 0.0780, 0.1840}
\definecolor{gray2}{rgb}{0.22, 0.22, 0.3}
\definecolor{ltgray2}{rgb}{0.647, 0.647, 0.647}
\definecolor{blueIV}{rgb}{0, 0, 0.7410}
\definecolor{blueIII}{rgb}{0.2, 0.2, 0.7410}
\definecolor{blueII}{rgb}{0.4, 0.4, 0.7410}
\definecolor{blueI}{rgb}{0.7410, 0.7410, 0.7410}
\definecolor{jetVI}{rgb}{0.9763 0.9831 0.0538}
\definecolor{jetV}{rgb}{0.9264 0.7256 0.2996}
\definecolor{jetIV}{rgb}{0.4783 0.7489 0.4877}
\definecolor{jetIII}{rgb}{0.0282 0.6663 0.7574}
\definecolor{jetII}{rgb}{0.0582 0.4677 0.8589}
\definecolor{jetI}{rgb}{0.2081 0.1663 0.5292}
\definecolor{mode1}{rgb}{0 0.4470 0.7410}
\definecolor{mode2}{rgb}{0.8500 0.3250 0.0980}
\definecolor{mode3}{rgb}{0.9290 0.6940 0.1250}
\definecolor{dkgold}{rgb}{0.5930 0.5150 0.3260}
\definecolor{mode1}{rgb}{0 0.4470 0.7410}
\definecolor{mode2}{rgb}{0.8500 0.3250 0.0980}
\definecolor{mode3}{rgb}{0.9290 0.6940 0.1250}
\definecolor{dkgold}{rgb}{0.5930 0.5150 0.3260}
\definecolor{matblue}{rgb}{0.000 0.447 0.741}
\definecolor{matred}{rgb}{0.850 0.325 0.098}
\definecolor{matyellow}{rgb}{0.9290 0.6940 0.125}
\definecolor{matpurple}{rgb}{0.494 0.184 0.556}
\definecolor{matgreen}{rgb}{0.466 0.674 0.188}
\definecolor{matcyan}{rgb}{0.3010 0.7450 0.9330}
\newcommand{\bs}{\boldsymbol}

\usepackage{hyperref}
\hypersetup{
	colorlinks = true,
	urlcolor = black,
	citecolor = blue2,
}

\title{Triglobal resolvent-analysis-based control of separated flows around low-aspect-ratio wings}

\author{
	{Jean H\'elder Marques Ribeiro\thanks{Corresponding author: jean.marques@fem.unicamp.br}   \ and Kunihiko Taira} \\
	Deptartment of Mechanical and Aerospace Engineeging\\ 
	University of California, Los Angeles \\
	Los Angeles, CA 90095, USA\\
}

\renewcommand{\shorttitle}{Triglobal resolvent-analysis-based flow control}

\begin{document}	
	\maketitle

	\begin{abstract}
		We perform direct numerical simulations (DNS) of actively controlled laminar separated wakes around low-aspect-ratio wings with two primary goals: (i) reducing the size of the separation bubble and (ii) attenuating the wing tip vortex. Instead of preventing separation, we modify the three-dimensional ($3$-D) dynamics to exploit wake vortices for aerodynamic enhancements. A direct wake modification is considered using optimal harmonic forcing modes from triglobal resolvent analysis. For this study, we consider wings at angles of attack of $14^\circ$ and $22^\circ$, taper ratios $0.27$ and $1$, and leading edge sweep angles of $0^\circ$ and $30^\circ$, at a mean-chord-based Reynolds number of $600$. The wakes behind these wings exhibit $3$-D reversed-flow bubble and large-scale vortical structures. For tapered swept wings, the diversity of wake vortices increases substantially, posing a challenge for flow control. To achieve the first control objective for an untapered unswept wing, root-based actuation at the shedding frequency is introduced to reduce the reversed-flow bubble size by taking advantage of the wake vortices to significantly enhance the aerodynamic performance of the wing. For both untapered and tapered swept wings, root-based actuation modifies the stalled flow, reduces the reversed-flow region, and enhances aerodynamic performance by increasing the root contribution to lift. For the goal of controlling the tip vortex, we demonstrate the effectiveness of actuation with high-frequency perturbations near the tip. This study shows how insights from resolvent analysis for unsteady actuation can enable global modification of $3$-D separated wakes and achieve improved aerodynamics of wings.
	\end{abstract}

	\section{Introduction}
	\label{sec:6_intro}
	Controlling $3$-D wake dynamics holds significant importance in aerospace engineering, especially in shaping the capabilities of future aircraft. One of the challenges in flow control involves managing massively separated $3$-D flows, as those observed behind finite-span wings at high angles of attack \citep{Anderson:10,Taira:JFM09}. Traditionally, aircraft operation under stalled flow conditions is avoided due to the detrimental effects associated with flow separation. Nevertheless, many studies have shown that biological flyers and swimmers benefit from stall under certain conditions \citep{Ellington:Nature96,Videler:Science04,Shyy08,Eldredge:ARFM19}. Moreover, for low-Reynolds-number flows over low-aspect-ratio wings, both lift and lift-to-drag ratio increase with the angle of attack within a certain range, even in the presence of a separation bubble \citep{Okamoto:AIAAJ2011aerodynamic,Ananda:AST15, MizoguchiTJSA:2016aerodynamic, Okamoto:TJSA2019disappearance}. This suggests potential benefits in exploring the flight operation of aircraft in stalled flow conditions once a flow control capable of enhancing  the aerodynamics of wings at high angles of attack~is~identified.
	
	Recent studies extensively delved into the fundamental characteristics of the separated wake around wings through experiments, computations, and theoretical analyses, primarily focusing on the low-Reynolds-number regime \citep{He:JFM17, Zhang:JFM20, Hayostek:AIAA22, Burtsev:JFM22, Ribeiro:JFM23triglobal, Pandi:JFM23}, thereby enhancing our understanding even for much higher-Reynolds-number flows. However, limited research has been conducted on controlling and modifying stalled flow behavior to enhance wing aerodynamics at high incidence angles \citep{Taira:AIAAJ09, Edstrand:JFM18b, Gopalakrishnan:AIAAJ17, Nastro:JFM23}.
	
	To improve the aerodynamic performance of wings, qualitatively evaluated as the lift-to-drag ratio, effective flow control strategies require a detailed study of load generation mechanisms. Given the particular case of stalled flows over low-aspect-ratio wings, a significant portion of lift originates from near-wake vortices around the wing \citep{Lee:JFM12,Zhang:JFM20, Zhang:PRF22, Ribeiro:JFM22,Ribeiro:JFM2023tapered}. However, due to the massive separation bubble over wings at high angles of attack, coherent structures associated with vortical lift tend to advect away from the wing, substantially reducing their contribution to the overall lift. 
	
	The reversed flow appears on the suction side of the wing for separated flows at high angles of attack \citep{Pauley:JFM90,Yarusevych:JFM09,Toppings:JFM22}.  To suppress this vortical formation and reattach the flow, boundary-layer control approaches using unsteady actuation exciting the shear layer instabilities can be efficient \citep{Seifert:JA96,Greenblatt:PAS00,Amitay:AIAAJ02}. For the low-Reynolds-number flow regime, suppression of the separation bubble impedes the formation of large-scale lift-related vortices around the wing. Moreover, identifying the optimal spatial-temporal characteristics of the control input to achieve a specific goal is nontrivial. Flow control effects depend on flow field characteristics, which are strongly associated with the wing planform geometry. 
	
	For untapered unswept wings, the reversed-flow bubble is larger inboard near the wing root, where spanwise-aligned vortex shedding structures emerge \citep{Taira:JFM09,Chen:EJM2016singletailed,Zhang:JFM20,Zhu:JFM2023swallow}. For swept wings, there is a spanwise shift of the reversed flow and shedding structures. This shift results from sweep-induced spanwise flow emerging over the wing and stabilizing wake oscillations  \citep{Ribeiro:JFM22,Burtsev:JFM22}. Over swept wings, the reversed-flow bubble spreads over a substantial portion of the wingspan, especially near the wing tip \citep{Zhang:JFM20b,Zhang:PRF22}. This large bubble formation makes it challenging to find an optimal location to perturb the reversed flow \citep{Mcfadden:AIAA22control,Brandt:AIAA23control}. 
	
	Despite differences in the wakes around unswept and swept wings at high incidence, they tend to exhibit similar lift-to-drag ratios. A higher lift and lift-to-drag ratio for laminar post-stall flow conditions, is achieved for tapered wings with high backward-swept leading edges \citep{Ribeiro:JFM2023tapered}. Wake patterns become increasingly complex around tapered wings, with vortex shedding spreading over a larger portion of the wingspan, while the wing tip vortex is considerably weakened compared to the tip vortex around an untapered wing.
	
	The tip vortex, a quasi-steady streamwise vortical structure emerging due to pressure differences between the upper and lower sides of the wing, is particularly interesting for laminar post-stall flows around untapered unswept wings \citep{Zhang:JFM20}. It can negatively affect the aerodynamics of wings, inducing drag, decreasing lift, and reducing the effective angle of attack near the tip \citep{Devenport:JFM96,Torres:AIAAJ04,Dong:EF20,Toppings:JFM21}. The emergence of the tip vortex and its interaction with the inboard wake vortices affects the aerodynamic loads over the wing and results in a complex $3$-D wake \citep{Freymuth:AIAAJ87,ViieruShyy:JA05,Visbal:AIAA12,Yilmaz:JFM12,Visbal:2013yilmazrockwell,Neal:PRF23,Zhu:JFM2024tip}. Additionally, its persistence can substantially impact air mobility and traffic control \citep{Spalart:ARFM98}. 
	
	The tip vortex is a common structure that appears over a broad range of Reynolds numbers. Control strategies that suppress its formation in the laminar regime provide insights for controlling tip vortices in turbulent flow conditions, which generally retain a laminar core. This is often observed for many other large-scale vortices emerging in post-stall wakes. Core wake structures in separated wakes exhibit topological similarities and find analogous vortical formations over a broad range of Reynolds numbers \citep{HuntEtAl:JFM1978,DallmannFDR1988,Delery:AR01,RibeiroBernardo:JFS20reynoldseffects}. This suggests that flow control findings for post-stall flows in the low-Reynolds-number regime can be valuable for higher-Reynolds-number flows. For the case of the tip vortex, its control and attenuation have benefited from the study of flow perturbations and instabilities \citep{Mayer:JFM92viscous,Edstrand:JFM18a,Edstrand:JFM18b}.
	
	Due to the heterogeneity of the $3$-D wake structures \citep{Zhang:JFM20b,Zhang:JFM20} and the nontrivial evolution of flow perturbations \citep{Navrose:JFM19}, the control of post-stall flows over wings may appear to be a daunting task. For such flows, an intuition-based control design may often lead to an ineffective flow modification and undesirable outcomes. To find strategies that efficiently modify the wakes, one may seek a proper spatial-temporal description of the actuation. This can be achieved studying the dynamics of optimal perturbations via triglobal resolvent analysis \citep{Ribeiro:JFM23triglobal,Ribeiro:AIAA23resolvent}.
	
	Resolvent analysis, one of the many techniques extracting important features from fluid flows \citep{Taira:AIAAJ17,Unnikrishnan:PAS23recent}, is particularly attractive for flow control as it identifies optimal inputs that can be amplified into the flow field   \citep{Trefethen:93,Jovanovic:JFM05,McKeon:JFM10}. Additionally, resolvent analysis reveals the unsteady response characteristics describing how optimal perturbations can potentially modify the base flow \citep{Luhar:JFM14opposition}. This method has been used to study a wide range of flow applications \citep{Moarref:JFM13,Schmidt:JFM18,Ricciardi:JFM22,Houtman:Flow23resolvent} and supported flow control designs over two-dimensional ($2$-D) base flows \citep{Yeh:JFM19,Jin:JFM2020feedbackresolvent,Liu:JFM21,Lin:JFM2023flow,Gross:AIAAJ2024resolventcontrol}. 
	
	\begin{figure}
		\footnotesize
		\centering
		\begin{tikzpicture}
		\node[anchor=south west,inner sep=0] (image) at (0,0) {\includegraphics[page=1,trim=4mm 0mm 4mm 0mm, clip,width=1\textwidth]{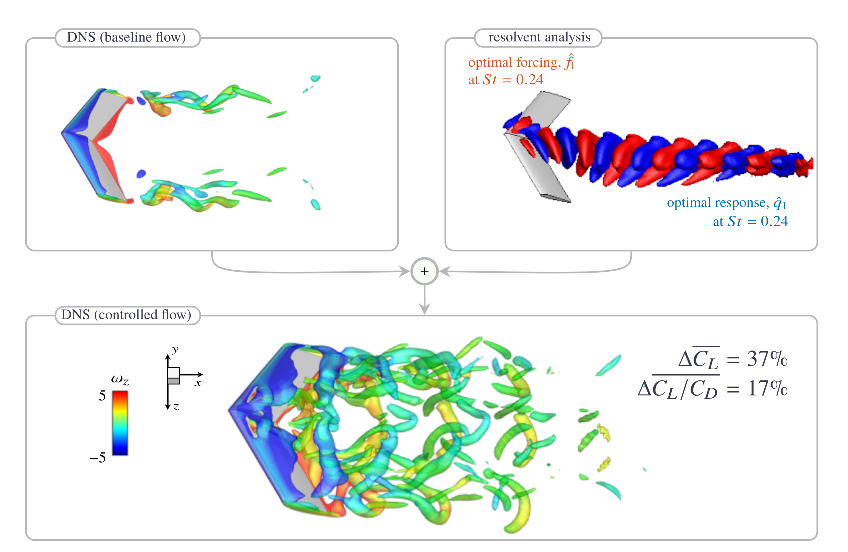}};
		\end{tikzpicture} \\ \vspace{-4mm}
		\caption{Overview of the present work shown for an example of a swept wing with $\Lambda = 30^\circ$ at $\alpha = 22^\circ$. DNS of baseline and controlled flows is visualized with isosurfaces of $Q = 1$ colored with spanwise vorticity $\omega_z$. Resolvent modes are visualized with isosurfaces of the $u_y$ component of forcing (top half span) $\hat{\mathbf{f}}_{uy} = \pm 1$ and response modes (bottom half span) $\hat{\mathbf{q}}_{uy} = \pm 0.5$. Full span wings are shown for visualization purposes. Optimal forcing modes at $St \equiv (\omega/2\pi)(c \sin \alpha / U_\infty \cos \Lambda) =  0.24$ are introduced to the baseline simulations as harmonic body forces. The actuation modifies the vortex shedding and yields $37\%$ lift and $17\%$ lift-to-drag ratio increase.} 
		\label{fig:6_schematics}
	\end{figure}

	Leveraging insights from the aforementioned studies, this work considers flow control over low-aspect-ratio wings with two primary objectives: (i) reducing the volume of the separation bubble to enhance time-averaged lift and lift-to-drag ratio and (ii) attenuating the wing tip vortex. To achieve these goals, we employ a direct wake modification approach \citep{Choi:ARFM08} based on resolvent analysis, by introducing triglobal forcing modes as harmonic body forces. Our intent is not to suppress the separation bubble and vortex shedding. Instead, our aim is to strategically modify the $3$-D wake dynamics to exploit vortices for aerodynamic enhancements.
		
	The overview of the present work is illustrated in figure \ref{fig:6_schematics}, where DNS flow structures are depicted using isosurfaces of the second invariant of the velocity gradient tensor, denoted as $Q$ \citep{Hunt:CTR1988eddies, Jeong:JFM95}, colored by the spanwise vorticity $\omega_z$, while forcing and response modes are depicted by the isosurfaces of the velocity component $u_y$. Our study is comprised of three parts: (i) obtain baseline flows (without control), (ii) obtain resolvent modes that reveals the most sensitive regions to introduce harmonic perturbations to the flow and their nonlinear response characteristics, and (iii) introduce the input (forcing) modes as harmonic body forces to the flows and analyze the effectiveness of the control strategy. Our work is organized as follows. In section \ref{sec:6_setup}, we describe our problem setup. In section \ref{sec:6_methods}, we present the numerical approach used in the present work; namely, DNS and triglobal resolvent analysis. In section \ref{sec:6_baseline}, we present the time-averaged characteristics and the vortex dynamics of the baseline flows. In section \ref{sec:6_resolvent}, the insights obtained from resolvent analysis and how it supports flow control.  In section \ref{sec:6_control}, we describe our active flow control approach and show its application to tapered swept wings and the effects of control. Our conclusions are presented in section~\ref{sec:6_conclusions}.
	
	\section{Problem setup}
	\label{sec:6_setup}
	
\begin{figure}
		\footnotesize
		\centering
		\begin{tikzpicture}
		\node[anchor=south west,inner sep=0] (image) at (0,0) {\includegraphics[page=1,trim=4mm 0mm 4mm 0mm, clip,width=1\textwidth]{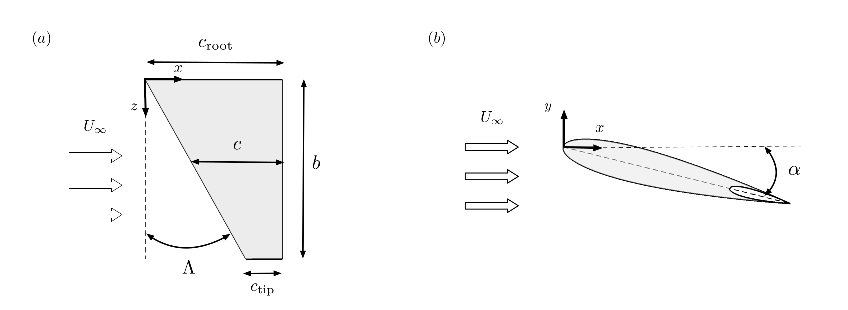}};
		\end{tikzpicture} \\ \vspace{-4mm}
		\caption{Setup for tapered swept wings. A wing with $sAR = 2$, $\Lambda = 30^\circ$ and $\lambda = 0.27$ is shown as an example in $(a)$ top and $(b)$ side view. The freestream velocity is shown by $U_\infty$. Geometry details are shown: $\Lambda$ is the sweep angle, $c$, $c_\text{tip}$, and $c_\text{root}$ are the mean, tip, and root chord lengths, respectively, $b$ is the half span, and $\alpha$ is the angle of attack.} 
		\label{fig:6_setup}
	\end{figure}

	We consider laminar flows over tapered swept wings with a NACA 0015 cross-sectional profile. For the Cartesian coordinate system, $(x,y,z)$ denote the streamwise, transverse, and spanwise directions, respectively. The origin is placed at the leading edge with the NACA 0015 profile prescribed along the $(x,y)$ plane. The angle of attack is defined between the airfoil chord line and the streamwise direction and set to $\alpha = 14^\circ$ and $22^\circ$ for this study. Our setup is detailed in figure \ref{fig:6_setup}. All wings considered herein present wing geometries with a sharp trailing edge and a straight-cut wing tip. 
	
	The $3$-D wing is built by extrusion from the wing root in the spanwise direction over a half-span length of $b$. Along with the mean chord length $c$, the semi aspect ratio of the wings is fixed as $sAR = b/c = 2$. Tapered wings consider a taper ratio defined by $\lambda = c_\text{tip} / c_\text{root}$, where $c_\text{tip}$ and $c_\text{root}$ are tip and root chord-lengths, respectively. We consider tapered wings with $\lambda = 0.27$ and $1$, where $\lambda = 1$ is the untapered wing. For swept wings, the $3$-D computational setup is sheared in the chordwise direction and $\Lambda$ is the sweep angle defined between the $z$-direction and the leading edge axis. We consider sweep angles $\Lambda = 0^\circ$ and $30^\circ$. 
	
	For all flows studied herein, the mean-chord based Reynolds number is defined as $Re_c \equiv U_\infty c /\nu = 600$, where $U_\infty$ is the freestream velocity, $c$ is the mean-chord length, and $\nu$ is the kinematic viscosity. Nondimensionalization is carried as follows: spatial variables are normalized by the mean-chord length $c$, velocities are normalized by the freestream velocity $U_\infty$, and time is reported in terms of convective time normalized by $c/U_\infty$.
	
	\section{Methodology}
	\label{sec:6_methods}
	
	\subsection{Direct numerical simulations}
	\label{sec:6_methods_dns}
	We study three-dimensional flows over wings by numerically solving the incompressible Navier--Stokes equations
	\begin{eqnarray}
	\frac{\partial \mathbf{u}}{\partial t} + \mathbf{u} \cdot \bs{\nabla} \mathbf{u} & = & -\bs{\nabla} p +\displaystyle{\frac{1}{Re_c}} \bs{\nabla}^2 \mathbf{u} + \mathbf{e} \mbox{ ,}\\
	\bs{\nabla} \cdot \bs{u} & = & 0 \mbox{ ,}
	\label{eq:6_NS}
	\end{eqnarray}
	where $\mathbf{u} = (u_x,u_y,u_z)$ is the velocity vector, $p$ is the pressure, and $\mathbf{e}$ is external forcing. The latter term is modeled as an harmonic body force using the spatial-temporal characteristics of the triglobal forcing modes obtained from resolvent analysis (section \ref{sec:6_methods_resolvent}). The set of equations \ref{eq:6_NS} is solved using \textit{Cliff}, the incompressible flow solver from the \textit{CharLES} package, developed by the Cascade Technologies, Inc. This solver uses collocated node-based second-order-accurate finite volume formulation to spatially discretize mass and momentum equations and a fractional step scheme for time integration \citep{Ham:CTR04energy,Ham:CTR06accurate}.
	
	With the origin of the Cartesian coordinate system placed at the leading edge of the wing root $(x,y,z)/c = (0,0,0)$, the computational domain extends over approximately $(x,y,z)/c \in [-20,25] \times [-20,20] \times [0,20]$. The present computational grids were also used in \cite{Ribeiro:JFM2023tapered}. A symmetry boundary condition is imposed at the root. At the inlet, we prescribe a freestream velocity vector $\bs{u} = (U_\infty,0,0)$. At the outlet, we specify the convective boundary condition. A slip boundary condition is set on all other farfield boundaries. Lastly, a no-slip wall boundary condition is enforced on the wing surface. 
	
	We perform DNS for baseline ($\bs{e} = 0$ in equation \ref{eq:6_NS}) and controlled flows ($\bs{e}$ modeled with harmonic forcing modes). For both cases, simulations are initiated from uniform flow with no external forcing, being performed with a constant Courant-Friedrichs-Lewy (CFL) number of 1 until transients are washed out of the computational domain, which takes approximately $t = 90$. After the transient flow features are washed out the domain, flows are simulated with and without external forcing using a constant time step defined such that CFL is smaller than one. Statistics are collected for approximately $t = 100$ to ensure statistical convergence.
	
	\subsection{Resolvent analysis}
	\label{sec:6_methods_resolvent}
\begin{figure}
		\footnotesize
		\centering
		\begin{tikzpicture}
		\node[anchor=south west,inner sep=0] (image) at (0,0) {\includegraphics[page=1,trim=4mm 0mm 4mm 0mm, clip,width=1\textwidth]{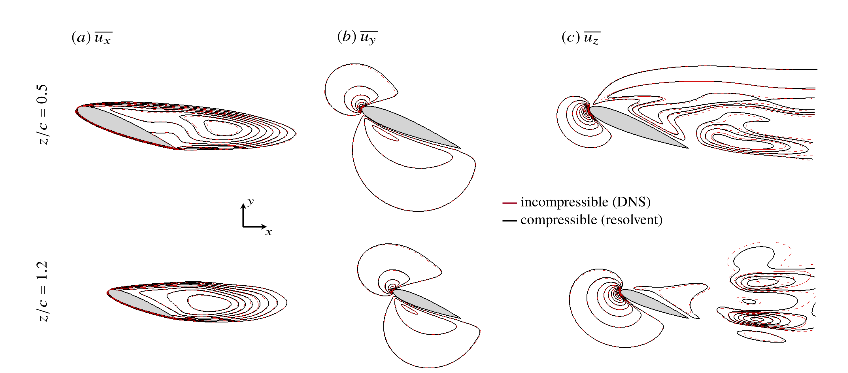}};
		\end{tikzpicture} \\ \vspace{-4mm}
		\caption{Time-averaged incompressible baseline flows from DNS (dashed, red) and the compressible base flows used for resolvent analysis (solid, black) shown as an example for the flow over a tapered swept wing with $\lambda = 0.27$ and $\Lambda = 30^\circ$ at $\alpha = 22^\circ$. Isocontour lines of $(a)$ $\overline{u_x} = [-0.3,0.3]$, $(b)$ $\overline{u_y} = [-0.3,0.7]$, and $(c)$ $\overline{u_z} = [-0.25,0.55]$ shown in $2$-D slices ($(x,y)$ plane) at $z/c = 0.5$ (top) and $1.2$ (bottom).} 
		\label{fig:6_icoComp}
	\end{figure}

	Resolvent analysis can provide valuable insights for the design of active flow control strategies. Let us consider the Reynolds decomposition of state variable $\mathbf{q} = \bar{\mathbf{q}} + \mathbf{q}^\prime$, where $\bar{\mathbf{q}}$ is the time-averaged flow and $\mathbf{q}'$ is the statistically stationary fluctuation component. In this study, the base flows $\bar{\mathbf{q}}$ used for resolvent analysis are derived from compressible flow simulations with a freestream Mach number $M_\infty \equiv U_\infty / a_\infty = 0.1$, where $a_\infty$ represents the freestream speed of sound. The spatial distribution of the velocity components for both incompressible and compressible time-averaged flows exhibits similar patterns, as depicted in figure \ref{fig:6_icoComp}. For compressible flows, the major variations in thermodynamic quantities occur closer to the wing surface and remain within a $1\%$ difference from the freestream values. Resolvent modes can be derived from both compressible and incompressible base flows. In the present work, a compressible resolvent approach is employed to build on our previous studies \citep{Ribeiro:JFM23triglobal,Ribeiro:AIAA23resolvent}.
	
	 With the compressible base flow, the aforementioned Reynolds decomposition is used to linearize the  compressible Navier--Stokes equations about $\bar{\bs{q}}$ to yield
	\begin{equation}
	\frac{\partial \mathbf{q}^\prime}{\partial t} = \mathbf{L}_{\bar{\mathbf{q}}}\mathbf{q}^\prime + \mathbf{f}^\prime \mbox{ ,}
	\label{eq:6_LNS}
	\end{equation}
	where $\mathbf{L}_{\bar{\mathbf{q}}}$ is the discrete form of the linearized Navier--Stokes operator and $\mathbf{f}^\prime$ encompasses the nonlinear and external forcing terms  \citep{McKeon:JFM10}. By considering the Fourier representation for $\mathbf{q}^\prime$ and $\mathbf{f}^\prime$, we have
	\begin{equation}
	\left[ \mathbf{q}^\prime(\mathbf{x},t), \mathbf{f}^\prime(\mathbf{x},t) \right] = \int_{-\infty}^{\infty} \left[ \hat{\mathbf{q}}_{\omega}(\mathbf{x}), \hat{\mathbf{f}}_{\omega}(\mathbf{x}) \right] e^{-i \omega t } {\text d} \omega \mbox{ ,}
	\label{eq:6_fourier}
	\end{equation}
	which transforms equation \ref{eq:6_LNS} to read
	\begin{equation}
	-i \omega \hat{\mathbf{q}}_{\omega} = \mathbf{L}_{\bar{\mathbf{q}}}\hat{\mathbf{q}}_{\omega} + \hat{\mathbf{f}}_{\omega} \mbox{ ,}
	\label{eq:6_LNS_fourier}
	\end{equation}
	where $\mathbf{x} = (x,y,z)$ and the triglobal response and forcing modes are $\hat{\mathbf{q}}_{\omega}$ and $\hat{\mathbf{f}}_{\omega}$, respectively, for a temporal frequency $\omega$. This expression leads to 
	\begin{equation}
	\hat{\mathbf{q}}_{\omega} = \mathbf{H}_{\bar{\mathbf{q}},\omega} \hat{\mathbf{f}}_{\omega} \mbox{ ,}
	\label{eq:6_cont_resolvent_1}
	\end{equation}
	with the resolvent operator $\mathbf{H}_{\bar{\mathbf{q}},\omega} \in \mathbb{C}^{m \times m}$. The operator size $m$ is defined by the product of the number of state variables and the number of spatial grid points. Herein, the linear operators have size $m$ of approximately $5 \times 10^6$. We analyze the resolvent operator through the singular value decomposition (SVD)
	\begin{equation}
	\mathbf{H}_{\bar{\mathbf{q}}} = \left[ -i \omega \mathbf{I} - \mathbf{L}_{\bar{\mathbf{q}}} \right]^{-1} = \mathbf{Q} \mathbf{\Sigma} \mathbf{F}^*,
	\label{eq:6_cont_resolvent_2}
	\end{equation}
	where $\mathbf{F} = [\hat{\mathbf{f}}_1,\hat{\mathbf{f}}_2,\dots,\hat{\mathbf{f}}_m]$ is an orthonormal matrix comprised of forcing modes, the diagonal matrix $\bs{\Sigma}=\text{ diag}[\sigma_1,\sigma_2,\dots,\sigma_m]$ holds the singular values (gain) in descending order, and $\mathbf{Q} = [\hat{\mathbf{q}}_1,\hat{\mathbf{q}}_2,\dots,\hat{\mathbf{q}}_m]$ is an orthonormal matrix comprised of response modes \citep{Trefethen:93,Jovanovic:JFM05}.
	
	The $\mathbf{H}_{\bar{\mathbf{q}},\omega}$ operators are discretized over $3$-D structured grids with the leading edge at the root positioned at $(x,y,z)/c = (0,0,0)$, extending over $(x, y, z)/c \in [-10,15] \times [-10,10] \times [0,10]$. The computational grids used for resolvent analysis are smaller in size than those used for DNS, we perform linear interpolation of the flow field between DNS and resolvent grids. Homogeneous Neumann boundary conditions are prescribed for $T'$ and homogeneous Dirichlet boundary conditions are set for the fluctuating variables $\rho'$ and $u'$ along the farfield, airfoil surface, and outlet. Far from the airfoil and in conjunction with the boundary conditions, sponges are applied \citep{Freund:AIAAJ97}.
	
	The resolvent modes are computed using the randomized resolvent analysis algorithm \citep{Ribeiro:PRF20} and the direct and adjoint linear systems were directly solved using the MUMPS (multifrontal massively parallel sparse direct solver) package \citep{Amestoy:SIAM01}. The adjoint-based sensitivity analysis was used to interpolate the resolvent norm over frequencies $\omega$ \citep{Schmid:AMR14,Fosas:JoT17}. The codes used to compute the resolvent modes are part of the \textit{linear analysis package} made available by \cite{SkeneRibeiro:tools}.
		
	\section{Baseline flows}
	\label{sec:6_baseline}
	\subsection{Time-averaged characteristics}
	\label{sec:6_baseline_timeaveraged}
\begin{figure}
		\footnotesize
		\centering
		\begin{tikzpicture}
		\node[anchor=south west,inner sep=0] (image) at (0,0) {\includegraphics[page=1,trim=4mm 0mm 4mm 0mm, clip,width=1\textwidth]{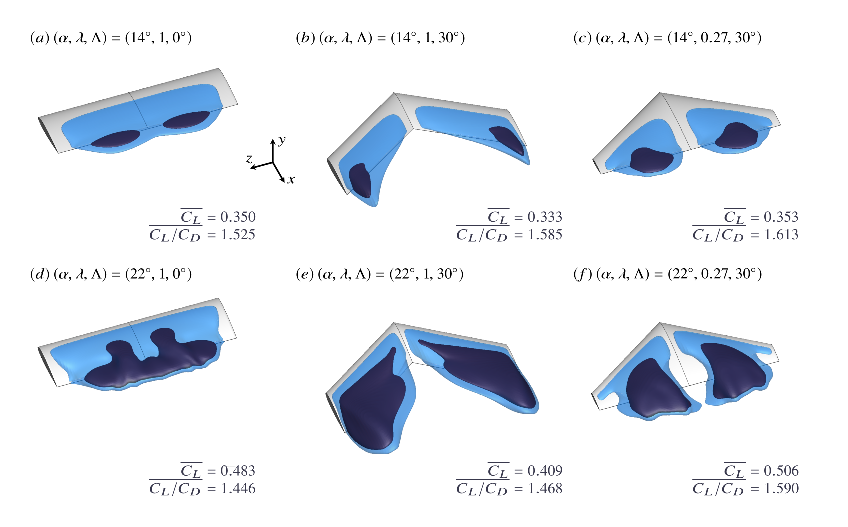}};
		\end{tikzpicture} \\ \vspace{-4mm}
		\caption{Time-averaged reversed flows over wings visualized with light blue and dark purple isosurfaces of $\overline{u_x} = 0$ and $-0.1$, respectively. Flow fields around wings with different $\Lambda$ and $\lambda$ combinations at ($a$-$c$) $\alpha = 14^\circ$ and ($d$-$f$) $22^\circ$ are presented. Wings are plotted with full span for visualization purposes. The time-averaged lift and lift-to-drag ratio, $\overline{C_L}$ and $\overline{C_L/C_D}$, respectively, are provided for each wing.} 
		\label{fig:6_separationBubble}
	\end{figure}

	To achieve the control objectives in this study, we first comprehensively examine the wake patterns and dynamics of the baseline flows. Specifically, for the control goal of reducing the separation bubble size, focus is placed on investigating key features of the reversed-flow bubble $3$-D topology through the time-averaged flow field. The reversed-flow bubble  has a nonhomogeneous spatial distribution, significantly influenced by both the wing planform geometry and the angle of attack, as illustrated in figure \ref{fig:6_separationBubble}. For visualization, we present the full span wing mirrored with respect to the wing root. The time-averaged reversed flow is depicted with the light blue and purple isosurfaces of $\overline{u_x} = 0$ and $-0.1$, respectively.  

	Additionally, we present the time-averaged aerodynamic forces, reported as non-dimensional drag and lift coefficients, ${C_D}$ and ${C_L}$ respectively, defined as
	\begin{equation}
	C_D = \frac{F_x}{\frac{1}{2}U_\infty^2 b c} \quad \text{and} \quad C_L = \frac{F_y}{\frac{1}{2}U_\infty^2 b c} \mbox{ ,}
	\label{eq:6_CL_CD}
	\end{equation} 
	where $F_x$ and $F_y$ are the $x$ and $y$ force components, respectively. For low-Reynolds-number flows, the time-averaged lift-to-drag ratio ($\overline{C_L/C_D}$), used throughout this work to quantify the aerodynamic performance of the wing, typically remains around  $\mathcal{O}(1)$ and reaches its peak at high angles of attack  \citep{Taira:JFM09,Zhang:JFM20}. Lift increases within the depicted $\alpha$ range, as seen in the $\overline{C_L}$ values in figures \ref{fig:6_separationBubble}($a$-$c$) for $\alpha = 14^\circ$ and \ref{fig:6_separationBubble}($d$-$f$) for $\alpha = 22^\circ$. Notwithstanding, the lift-to-drag ratio decreases with $\alpha$, indicating loss of aerodynamic performance as the angle of attack increases.  This decline in aerodynamic efficiency at $\alpha = 22^\circ$, compared to $\alpha = 14^\circ$, is linked to a post-stall flow condition. Consequently, a flow control implementation becomes essential to assist the wing in maintaining aerodynamic efficiency. 
	
	Wing sweep and taper also influence aerodynamic loads. Overall, wing sweep is associated with a reduction in lift for untapered wings \citep{Zhang:JFM20b}. The combination of a high leading edge sweep and wing taper proves advantageous for the aerodynamic performance of the wing, especially for post-stall flows \citep{Ribeiro:JFM2023tapered}. Achieving a high lift-to-drag ratio is possible for steady wakes at lower angles of attack. However, for enhanced aerodynamic performance, the role of flow unsteadiness must be considered \citep{Zhang:JFM20}. To take advantage of the inherent unsteadiness, we consider resolvent-analysis-based actuation that can excite unsteady structures in the wake. For realizing effective flow modification, we first investigate the characteristics of the reversed-flow bubble around the wings shown in figure \ref{fig:6_separationBubble}.
	
	For laminar flows over wings at high incidence angles, the separation bubble becomes massive, beyond the size of the wing. As shown in figure \ref{fig:6_separationBubble}, the size of the reversed-flow bubble is more pronounced for flows over wings at $\alpha = 22^\circ$ and the intensity of the reversed flow within the separation bubble increases with the angle of attack, as evident from the larger $\overline{u_x} = -0.1$ isosurfaces. Additionally, the nonhomogeneity of the reversed flow is observed in the spanwise direction, with the bubble expanding and contracting over the wingspan. For example, for untapered unswept wings $(\lambda,\Lambda = 1,0^\circ)$, the bubble shrinks near the wing tip while expanding in the inboard region.
	
	The wingspan location where the separation bubble reaches its maximum sectional area in the $(x,y)$ plane is a key aspect of the reversed-flow topology. For control, this location is important as a direct wake modification implemented over this region can significantly reduce the reversed-flow volume. For untapered unswept wings at lower $\alpha$, the reversed-flow bubble is marked by the contraction of the reversed-flow region near the midspan, as depicted in figure \ref{fig:6_separationBubble}$(a)$. \cite{Zhu:JFM2023swallow} noted that similar reversed-flow bubble shapes resemble the tail of a swallow and termed this reversed-flow topology as a swallow-tailed structure. With an increase in the angle of attack, they noted that the bubble transforms into a single-tailed formation, where the tail refers to portion of the reversed-flow bubble with larger extension downstream of the wing in the streamwise~direction. A single-tailed reversed-flow structure is analogous to the structure presented in figure \ref{fig:6_separationBubble}$(d)$. Such wake formation is commonly observed for $3$-D flows over untapered unswept low-aspect-ratio wings at high angles of attack \citep{Chen:EJM2016singletailed}. 
	
	In contrast to the flows over untapered unswept wings, for swept wings $(\Lambda = 30^\circ)$ the flow is attached over the midspan. For untapered swept wings  $(\lambda,\Lambda = 1,30^\circ)$, shown in figure \ref{fig:6_separationBubble}$(b,e)$, the larger portion of the reversed-flow bubble emerges over the outboard side of the wing, closer to the tip. The characteristics of the reversed-flow bubble are further influenced by wing taper, as shown in figure \ref{fig:6_separationBubble}$(c,f)$. For tapered swept wings $(\lambda,\Lambda = 0.27,30^\circ)$, the larger portion of the reversed-flow structure appears over the inboard section of the wing. To gain further insights into the wakes, we examine the vortex dynamics of the wakes in the following~section.
	
	\subsection{Vortex dynamics}
	\label{sec:6_baseline_vortexdynamics}
	\begin{figure}
	\footnotesize
	\centering
	\begin{tikzpicture}
		\node[anchor=south west,inner sep=0] (image) at (0,0) {\includegraphics[page=1,trim=4mm 0mm 4mm 0mm, clip,width=1\textwidth]{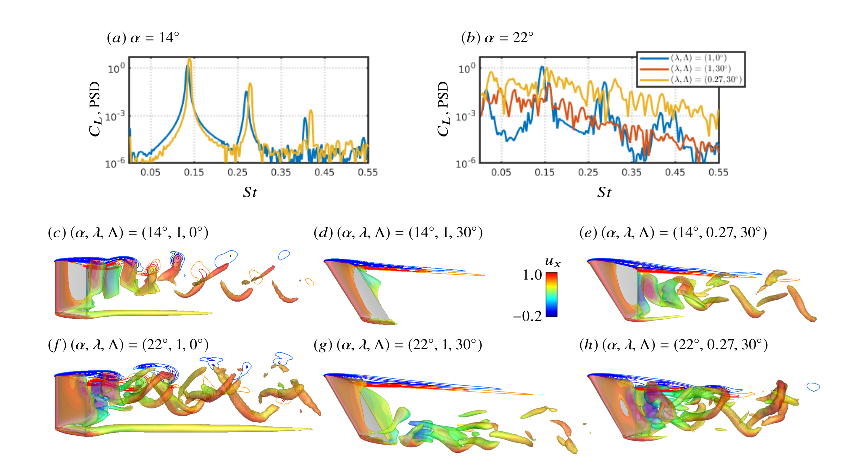}};
	\end{tikzpicture} \\ \vspace{-4mm}
	\caption{$(a,b)$ Power spectrum density (PSD) of $C_L$ and ($c$-$h$)nstantaneous flow fields around tapered swept wings visualized with isosurfaces of $Q = 1$ colored by the streamwise velocity $u_x$. Along the root plane, spanwise vorticity $\omega_z$ contours are shown. We present flow fields around wings with different $\Lambda$ and $\lambda$ combinations at ($c$-$e$) $\alpha = 14^\circ$ and ($f$-$h$) $22^\circ$.} 
	\label{fig:6_dns}
	\end{figure}
	
	Let us continue to study the features of flows over low-aspect-ratio wings by examining  their spectral characteristics and instantaneous wakes illustrated in figure \ref{fig:6_dns}. In figures \ref{fig:6_dns}$(a,b)$, we present the power spectra density (PSD) of $C_L$, where frequencies are normalized as
	\begin{equation}
	St \equiv \frac{\omega}{2\pi} \frac{c \sin \alpha}{ U_\infty \cos \Lambda} \mbox{   ,}
	\label{eq:strouhal}
	\end{equation}
	which is a modified Fage--Johansen Strouhal number \citep{FageJohanssen:PRSA27} with a $1/\cos \Lambda$ scaling to account for the sweep angle. The lift spectra reveal peaks associated with vortex shedding. At $\alpha = 14^\circ$, the shedding structures are predominantly $2$-D near the wing root, resulting in smoother lift PSD curves when compared to the ones obtained at $\alpha = 22^\circ$, where shedding vortices are $3$-D.  Despite the differences in wake shedding, lift peaks are found between $0.13 \le St \le 0.16$ for all cases shown. Within this frequency range, untapered unswept wings $(\lambda,\Lambda = 1,0^\circ)$ peak around the lower frequencies $St \approx 0.14$ while tapered swept wings $(\lambda,\Lambda = 0.27,30^\circ)$ peak around the higher portion of the frequency range $St \approx 0.16$. In figures \ref{fig:6_dns}$(c,f)$, we examine the time-varying flows over low-aspect-ratio untapered unswept wings $(\lambda,\Lambda = 1,0^\circ)$ at $\alpha = 14^\circ$ and $22^\circ$. Notably, the global unsteady characteristics of the flow share similarities for both angles of attack. Near the root, unsteadiness is predominantly marked by the emergence of vortices aligned in the spanwise direction. At the wing tip, a steady flow displays a tip vortex oriented along the streamwise direction. The streamwise length of this tip vortex is more pronounced for $\alpha = 22^\circ$.
	
	Wing sweep plays a stabilizing role in attenuating wake oscillations around $3$-D wings \citep{Zhang:JFM20b, Burtsev:JFM22}, as shown in figures \ref{fig:6_dns}$(d,g)$. At the lower $\alpha$, unsteadiness is entirely suppressed. However, at $\alpha=22^\circ$, the flow exhibits unsteadiness near the wing tip, while remaining steady near the wing root. For the higher $\alpha$, the unsteady flow structures develop farther away from the wing. 
	
	Figures \ref{fig:6_dns}$(e,h)$ reveal that the wake patterns of tapered swept wings $(\lambda,\Lambda = 0.27,30^\circ)$ exhibit notable similarities with those around untapered unswept wings $(\lambda,\Lambda = 1,0^\circ)$. However, a distinctive feature is the significant shortening of the tip vortex in tapered swept wings. Despite the pronounced leading edge sweep angle, the wakes of tapered swept wings are unsteady. Vortex shedding structures feature spanwise rolls, akin to the flows around untapered unswept wings shown in figures \ref{fig:6_dns}$(c,f)$. The shift in flow unsteadiness along the spanwise direction, particularly toward the quarter-span (located halfway between wing the root and tip), coincides with the larger expansion of the reversed-flow bubble size over the wingspan, as depicted in figure \ref{fig:6_separationBubble}$(e,h)$.
	
	\section{Triglobal resolvent analysis}
	\label{sec:6_resolvent}
\begin{figure}
		\footnotesize
		\centering
		\begin{tikzpicture}
		\node[anchor=south west,inner sep=0] (image) at (0,0) {\includegraphics[page=1,trim=4mm 0mm 4mm 0mm, clip,width=1.0\textwidth]{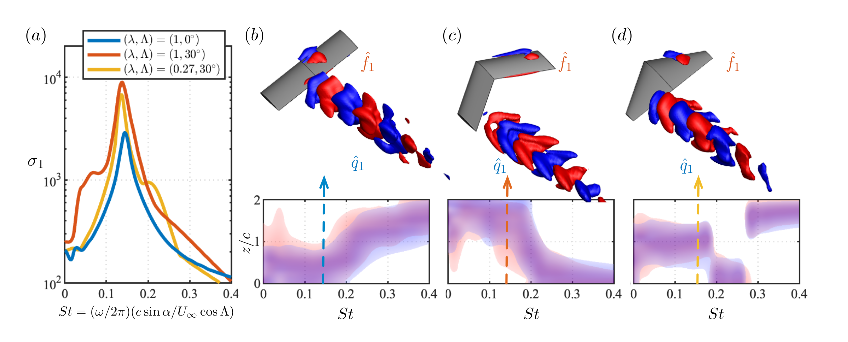}};
		\end{tikzpicture} \\ \vspace{-4mm}
		\caption{Resolvent modes for flows around low-aspect-ratio wings at $\alpha = 22^\circ$. $(a)$ Leading amplification gain $\sigma_1$ spectra. $(b,c,d)$ Bottom, heatmap of the spatial support of forcing (red) and response (blue) modes at each $St$, visualized by  $\Omega_{\hat{\mathbf{f}}}/\| \Omega_{\hat{\mathbf{f}}} \|_\infty \ge 0.5$ and $\Omega_{\hat{\mathbf{q}}}/\| \Omega_{\hat{\mathbf{q}}} \|_\infty \ge 0.5$, respectively. An arrow marks the $St$ of the mode shown on top. Isosurfaces of forcing $\hat{\mathbf{f}}_{uy} = \pm 1$ and response modes $\hat{\mathbf{q}}_{uy} = \pm 0.5$ are shown for $(b)$ $(\lambda,\Lambda) = (1,0^\circ)$, $(c)$ $(\lambda,\Lambda) = (1,30^\circ)$, and $(d)$ $(\lambda,\Lambda) = (0.27,30^\circ)$ wings. } 
		\label{fig:6_resolvent}
	\end{figure}

	Resolvent analysis reveals the spatial regions with high receptivity to external perturbations, the optimal frequency of actuation, and the harmonic response of the flow field. Those insights can support flow control design by uncovering the responsiveness of the flow field to optimal perturbations. We discuss how the findings obtained from resolvent analysis can support flow control using the resolvent modes over wings at $\alpha = 22^\circ$, as shown in figure \ref{fig:6_resolvent}.  We present the leading resolvent gain spectra and the forcing-response mode pairs at $St = 0.14$. As seen in figure \ref{fig:6_resolvent}$(a)$, the frequency peaks for the highest gain $\sigma_1$ at $St \approx 0.14$ are consistent across various wing planforms. Another commonality in the resolvent modes is attributed to the convective nature of the resolvent operator, where forcing modes manifest upstream of the wing, while response modes emerge downstream in the wake.
	
	Resolvent modal structures exhibit a pattern of spatial oscillations in the streamwise direction, whose wavelength scale relates to the modal frequency and the convective physics of the base flow. For this reason, lower-frequency resolvent modes usually have larger streamwise wavelengths, while higher-frequency resolvent modes exhibit shorter streamwise wavelengths. It is important to note that resolvent modes are defined over the entire domain, albeit being in a non-uniform manner. That is, resolvent mode have a localized larger amplitude over certain regions, i.e., regions with higher spatial support of forcing or response modes. For the flow fields considered herein, the higher spanwise support of each pair of forcing and response modes appears over the same portion of the wingspan. As the frequency increases, there is a shift in the spatial support of forcing-mode pairs across the wingspan.
	
	We illustrate the connection between the spatial support of forcing-response mode pairs over the wingspan $(z/c)$ and their frequency $(St)$ in figures \ref{fig:6_resolvent}($b$-$d$) using the contours of
	\begin{equation}
	\Omega_{\hat{\mathbf{f}}}(z) = \int_{S(x,y)} \|\hat{\mathbf{f}}\|_2 \ {\rm d} S \quad \text{and} \quad \Omega_{\hat{\mathbf{q}}}(z) = \int_{S(x,y)} \|\hat{\mathbf{q}}\|_2 \ {\rm d} S \mbox{   ,}
	\label{eq:integralF}
	\end{equation}
	where $\|\hat{\mathbf{f}}\|_2$ and $\|\hat{\mathbf{q}}\|_2$ are the $2$-norm of $\hat{\mathbf{f}}$ and $\hat{\mathbf{q}}$, respectively, at each grid point of the computational domain. These modes are defined by five state variables, denoted as  $\hat{\mathbf{f}} = [\hat{\mathbf{f}}_\rho,\hat{\mathbf{f}}_{ux},\hat{\mathbf{f}}_{uy},\hat{\mathbf{f}}_{uz},\hat{\mathbf{f}}_T]$ and $\hat{\mathbf{q}} = [\hat{\mathbf{q}}_\rho,\hat{\mathbf{q}}_{u_x},\hat{\mathbf{q}}_{u_y},\hat{\mathbf{q}}_{u_z},\hat{\mathbf{q}}_T]$, under the Chu norm \citep{Chu:Acta65}. As $\|\hat{\mathbf{f}}\|_2$ and $\|\hat{\mathbf{q}}\|_2$ are integrated over spanwise slices of the domain, $S(x,y)$, the $\Omega_{\hat{\mathbf{f}}}$ and $\Omega_{\hat{\mathbf{q}}}$ shows the wingspan region over which the forcing-response mode pair has higher spatial support. For the visualization in figure \ref{fig:6_resolvent}($b$-$d$), $\Omega_{\hat{\mathbf{f}}}$ and $\Omega_{\hat{\mathbf{q}}}$ are normalized based on their respective maximum values at each $St$. These contours are then represented using red and blue shades, where $\Omega_{\hat{\mathbf{f}}}/\| \Omega_{\hat{\mathbf{f}}} \|_\infty \ge 0.5$ and $\Omega_{\hat{\mathbf{q}}}/\| \Omega_{\hat{\mathbf{q}}} \|_\infty \ge 0.5$, respectively. This approach results in a heatmap illustrating the spatial distribution of  $\hat{\mathbf{f}}$ and $\hat{\mathbf{q}}$ across the wingspan for each frequency.

	The spatial support over the wingspan for the forcing-response mode pairs at the frequency corresponding to the resolvent gain peak coincides with the spanwise location where the reversed-flow bubble is larger.  Specifically for the untapered unswept wing $(\lambda,\Lambda = 1,0^\circ)$ at $\alpha = 22^\circ$, the dominant mode pair at $St = 0.14$ displays root-dominant modal structures seen in figure \ref{fig:6_resolvent}$(b)$ analogous to the shape of the shedding vortices observed in the DNS. However, with increasing frequency, the spatial support of the forcing-response mode pairs gradually transitions in the spanwise direction from the root region towards the wing tip. Resolvent analysis indicates that lower frequency actuation ($St \le 0.20$) is more responsive and prone to altering wake dynamics when applied near the wing root. Conversely, actuation at $St > 0.20$ demonstrates greater potential for modifying the wake closer to the wing~tip.
	
	With the insights from the resolvent modes, we can strategically determine the region within the flow field that amplifies unsteadiness to attain a specific control objective. For the untapered swept wing illustrated in figure \ref{fig:6_resolvent}$(c)$, the larger portion of the reversed-flow bubble appears at the wing tip. Consequently, the forcing-response mode pairs at the peak resolvent gain frequency, $St = 0.14$, emerge closer to the wing tip region. As frequencies increase, spatial support shifts from the wing tip towards the wing root for $St > 0.20$. 
	
	For the flow over the tapered swept wing $(\lambda,\Lambda = 0.27,30^\circ)$ depicted in figure \ref{fig:6_resolvent}$(d)$ the reversed-flow bubble is prominent over the quarter-span. Therefore, the forcing-response mode pairs at the resolvent gain frequency peak also emerge near the quarter-span. There is a distinct shift of spatial support towards the wing root for $0.20 \le St < 0.30$, and at higher frequencies ($St \ge 0.30$), the modes shift towards the wing tip. For the sake of brevity, we do not discuss the suboptimal resolvent modes, since we are interested in flow control techniques based on the primary modes. Discussion on the subdominant modes can be found in \cite{Ribeiro:JFM23triglobal,Ribeiro:AIAA23resolvent}.
		
	\section{Controlled flows}
	\label{sec:6_control}
	
	\subsection{Implementation of resolvent-analysis-based actuation}
	\label{sec:6_methods_resolvent_actuation}
	Resolvent analysis finds the optimal perturbation that can be amplified in the flow field.  However, the receptivity of low-Reynolds-number separated $3$-D flows to external perturbations remains elusive. To address this issue, forcing modes obtained from resolvent analysis are introduced as an harmonic external body force $\bs{e}$ in equation \ref{eq:6_NS}. The body force is used in the numerical simulation to model actuators employed in practical flow control applications. For instance, the control effects of dielectric barrier discharge (DBD) plasma actuators have been studied numerically using body forces \citep{Shyy:JAP2002bodyforce,Mullenix:AIAAJ2013bodyforce,WaindimGaitonde:JCP2016bodyforce} and surface heat actuators \citep{Yeh:JFM2017laminar,Prasad:JFM2023plasma}. For this reason, even though this study is a fundamental analysis of the $3$-D post-stall flow control, the spatio-temporal characteristics of the body forces presented here can be approximated in practical flow control applications. 
	
	The spatial distribution of the forcing modes and their frequency $\omega$ is used within the definition of the body force $\bs{e}$ expressed as
	\begin{equation}
	\bs{e}(\bs{x},t) = A [\text{Re}(\hat{\bs{f}}(\bs{x}))\sin(\omega t + \phi) + \text{Im}(\hat{\bs{f}}(\bs{x}))\cos(\omega t + \phi)] \mbox{ ,}
	\label{eq:6_BF}
	\end{equation}
	where $\hat{\bs{f}}(\bs{x})$ is the spatial forcing mode, with real $\text{Re}(\cdot)$ and imaginary $\text{Im}(\cdot)$ parts, $A$ is the amplitude, and $\phi$ is the phase. For all controlled flows considered herein, $\phi$ is set to initiate the body force actuation at the time of maximum $C_L$. The resolvent modes $\hat{\mathbf{f}}$ and $\hat{\mathbf{q}}$ are weakly compressible, being derived from a base flow characterized by a freestream Mach number of $M_\infty = 0.1$ \citep{Ribeiro:AIAA23resolvent}. Only velocity components of $\hat{\mathbf{f}}$ are used in equation \ref{eq:6_BF}.
	
	\begin{figure}
		\footnotesize
		\centering
		\begin{tikzpicture}
		\node[anchor=south west,inner sep=0] (image) at (0,0) {\includegraphics[page=1,trim=4mm 4mm 9mm 0mm, clip,width=1\textwidth]{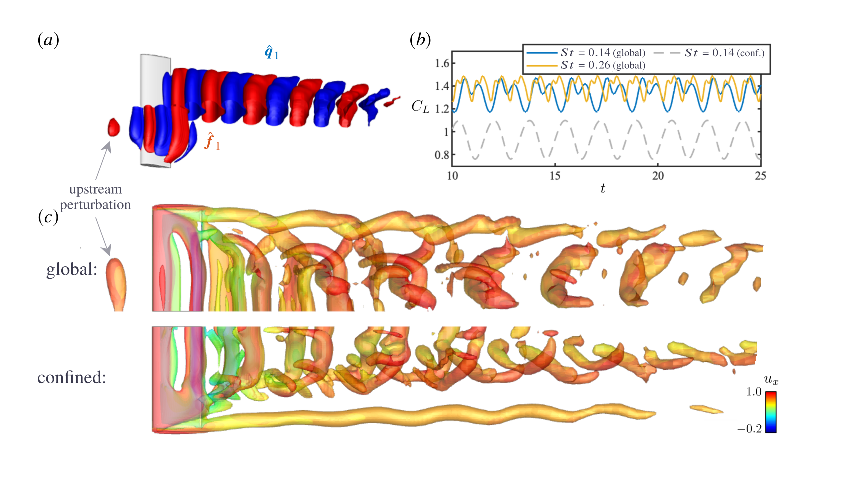}};
		\end{tikzpicture} \\ \vspace{-4mm}
		\caption{Global and confined forcing mode actuation for the flow control over a untapered unswept wing $(\lambda,\Lambda = 1,0^\circ)$ at $\alpha = 14^\circ$. $(a)$ Isosurfaces of $St = 0.14$ forcing $\hat{\mathbf{f}}_{uy} = \pm 0.5$ (bottom) and response $\hat{\mathbf{f}}_{uy} = \pm 0.2$ (top). $(b)$ $C_L$ over time for global and confined controlled flows at $2$ actuation frequencies. $(c)$ Global and confined controlled flows using $St = 0.14$ forcing mode, visualized with $Q = 0.3$, colored by $u_x$. Arrows show how upstream perturbation emerges in the forcing mode and how it appears in the flow actuated using global forcing.} 
		\label{fig:6_local}
	\end{figure}
		
	Forcing modes are spatially global, extending over the entire computational domain $(\bs{x})$, with higher spatial support upstream of the wing, as depicted in figure \ref{fig:6_local}$(a)$.  To provide fundamental insights into separation control, we narrow our focus to actuation over a specific subset of $\bs{x}$. We term this approach as spatially confined actuation, in which body forces are introduced within a volume of actuation ($V_\text{act}$) and its corresponding actuation surface ($S_\text{act}$), defined as the outer surface of the actuation volume. Our procedure to define $V_\text{act}$ will be discussed in the following section. It is noteworthy that global and localized actuation lead to distinct control outcomes, as depicted in figure \ref{fig:6_local}$(b,c)$. 
		
	This study adopts a confined actuation approach for two primary reasons. First, global forcing modes emerge upstream far from the wing, making them unrealistic for practical aircraft actuation and flow control.	Secondly, confined actuation enables us to discern the effects of control over specific regions of the wingspan. The spanwise support of forcing modes changes as the frequency of actuation varies (section \ref{sec:6_resolvent}). Let us take the example of a global forcing actuation using forcing modes at $St = 0.14$ and $0.26$ using the same body force amplitude $A$. Notably, the forcing modes at $St = 0.14$ have stronger spatial support over the wing root, while those at $St = 0.26$ are more pronounced at the wing tip. Nevertheless, global actuation yields similar lift over time for both frequencies, as seen in figure \ref{fig:6_local}$(b)$. Additionally, global actuation with the $St = 0.14$ forcing mode alters the wake across the entire wingspan, perturbing the tip vortex upstream in the wake, as observed in figure \ref{fig:6_local}($c$). In contrast, the control effects of the confined actuation in the wake are concentrated in~the~root~region.
	
	\subsection{Volume of actuation and momentum coefficient}
	\label{sec:6_methods__vact}
	The objective of control is to achieve notable flow modifications using a small forcing input. To quantify the magnitude of the control input, we consider the momentum coefficient $C_\mu$ defined as
	\begin{equation}
	C_\mu \equiv \frac{\rho\frac{S_\text{act}}{V_\text{act}}\int_{V_\text{act}}\left( {u_x^\prime}^2 + {u_y^\prime}^2 + {u_z^\prime}^2 \right) \text{d}V}{\frac{1}{2}\rho U_\infty^2 bc} \mbox{ ,}
	\label{eq:6_cmu}
	\end{equation}
	where ${u_x^\prime}$, ${u_y^\prime}$, and ${u_z^\prime}$ are the velocity components induced by the actuation body force $\bs{e}$ defined in $V_\text{act}$. The induced velocity is measured in a calibration simulation in which the freestream flow is turned off. The determination of an actuation volume involves the analysis of the magnitude of $\hat{\mathbf{f}}$:
	\begin{equation}
	K_{\hat{\mathbf{f}}}(\mathbf{x},\omega) \equiv | \hat{\mathbf{f}}_{u_x}^*\hat{\mathbf{f}}_{u_x} + \hat{\mathbf{f}}_{u_y}^*\hat{\mathbf{f}}_{u_y} + \hat{\mathbf{f}}_{u_z}^*\hat{\mathbf{f}}_{u_z} | \mbox{ .}
	\label{eq:6_kineticForcing}
	\end{equation}
	This scalar variable is defined across the spatial domain for each forcing mode and serves as a velocity-based metric depicting the spatial support of the forcing modes. 
	
\begin{figure}
		\footnotesize
		\centering
		\begin{tikzpicture}
		\node[anchor=south west,inner sep=0] (image) at (0,0) {\includegraphics[page=1,trim=4mm 0mm 1mm 0mm, clip,width=1\textwidth]{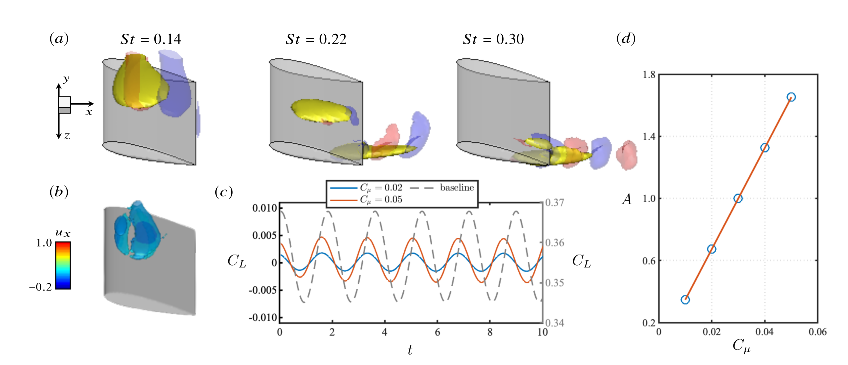}};
		\end{tikzpicture} \\ \vspace{-4mm}
		\caption{Confined forcing and amplitude definition for untapered unswept wing $(\lambda,\Lambda = 1,0^\circ)$ at $\alpha = 14^\circ$. $(a)$ Actuation volume $V_\text{act}$ (yellow) and forcing modes visualized with isosurfaces of $\hat{\mathbf{f}}_{ux}/\| \hat{\mathbf{f}}_{ux} \|_\infty = \pm 0.1$ (blue-red). $(b)$ Actuation in quiescent flow visualized with isosurfaces of $Q = 0.1$ colored by $u_x$ for $St = 0.14$ forcing modes. $(c)$ Lift coefficient $C_L$ over time for baseline (right axis) and actuated quiescent flows (left axis) $(d)$ Amplitude of body force actuation $A$ for estimated momentum coefficients of $0.01 \le C_\mu \le 0.05$.} 
		\label{fig:6_forcingLocation}
	\end{figure}

	The actuation volume, denoted as $V_\text{act}$, is specified within the region where
	\begin{equation}
	K_{\hat{\mathbf{f}}}(\mathbf{x},\omega) \ge 0.5 \max_{\mathbf{x}}(K_{\hat{\mathbf{f}}}(\mathbf{x},\omega))
	\label{eq:6_Kmax} \mbox{ .}
	\end{equation} 
	This threshold is selected for two main reasons: (i) $V_\text{act}$ is limited to a region closer to the wing surface for all cases studied herein and (ii) the region of spanwise support for the forcing modes can be clearly distinguished for modes at different frequencies (see figure \ref{fig:6_resolvent}). A yellow isosurface illustrates the definition of $V_\text{act}$, as seen in  figure \ref{fig:6_forcingLocation}, along with red and blue isosurfaces representing the global modes with $\hat{\mathbf{f}}_{ux} / \| \hat{\mathbf{f}}_{ux} \|_\infty = \pm 0.1$.
	
	Following the definition of $V_\text{act}$, simulations are conducted under quiescent freestream flow with actuation to estimate the momentum coefficient $C_\mu$. The velocity components of the forcing mode, denoted as $\hat{\mathbf{f}} = [\hat{\mathbf{f}}_{ux},\hat{\mathbf{f}}_{uy},\hat{\mathbf{f}}_{uz}]$, each having unit magnitude ($\| \hat{\mathbf{f}} \|_2 = 1$), are introduced to the momentum equation \ref{eq:6_NS}, as illustrated in figure \ref{fig:6_forcingLocation}$(b)$, with $Q = 0.1$. The chosen value of $Q$ in this context is $10$ times lower than that used throughout the manuscript, to enlarge flow structures for visualization purposes. As shown in figure \ref{fig:6_forcingLocation}$(c)$, lift fluctuation induced by actuation under quiescent freestream flow conditions remains under approximately $1\%$ of the baseline $C_L$ for $C_\mu \le 0.05$. Through simulations of quiescent freestream flows, we obtain $u_x^\prime$, $u_y^\prime$ and $u_z^\prime$ for equation \ref{eq:6_cmu} to be evaluated over $V_\text{act}$, revealing a linear relation between $C_\mu$ and the actuation amplitude $A$, as illustrated in figure \ref{fig:6_forcingLocation}$(d)$. For $C_\mu = 0.05$, the value of $A$ is of order $\mathcal{O}(1)$, exhibiting a magnitude similar to previous studies implementing body-force actuation in comparable flow conditions \citep{Edstrand:JFM18b}.
	
\begin{figure}
		\footnotesize
		\centering
		\begin{tikzpicture}
		\node[anchor=south west,inner sep=0] (image) at (0,0) {\includegraphics[page=1,trim=10mm 0mm 10mm 0mm, clip,width=1.0\textwidth]{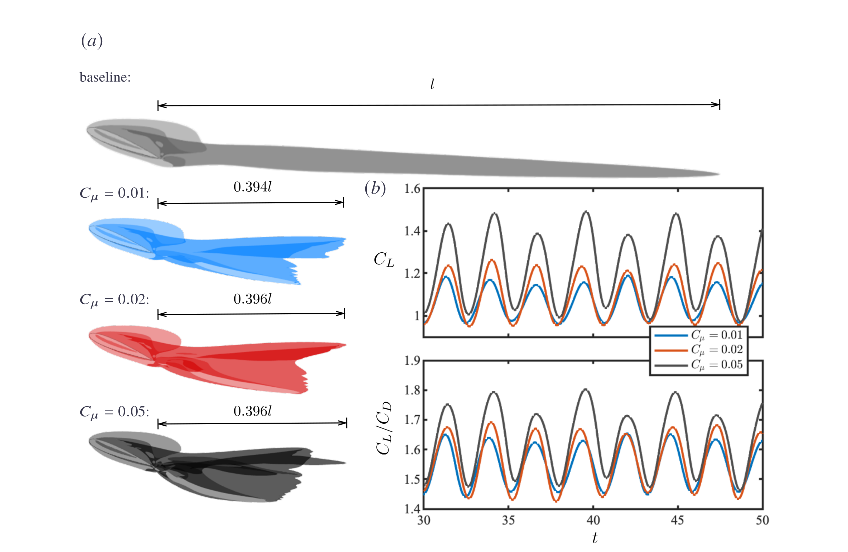}};
		\end{tikzpicture} \\ \vspace{-4mm}
		\caption{Effect of momentum coefficient $C_\mu$ on controlled flows over an untapered unswept wing $(\lambda,\Lambda = 1,0^\circ)$ at $\alpha = 22^\circ$ using $(a)$ $St = 0.40$ (wing tip) and $(b)$ $St = 0.14$ (reversed flow) forcing modes. $(a)$ Tip vortex attenuation visualized with time-averaged isosurfaces of $\overline{Q} = 1$ colored in light gray for baseline, blue for $C_\mu = 0.01$, red for $C_\mu = 0.02$, and dark gray for $C_\mu = 0.05$. $(b)$ $\overline{C_L}$ and $\overline{C_L/C_D}$ over time for controlled flows with distinct input~$C_\mu$.} 
		\label{fig:6_cmuEffects}
	\end{figure}

	In our analysis, we maintain a constant $C_\mu = 0.05$. It is noteworthy that variations in $C_\mu$ may induce changes in the controlled wake, as illustrated in figure \ref{fig:6_cmuEffects}. This figure depicts the impact of different $C_\mu$ values on flow control over an untapered unswept wing $(\lambda,\Lambda = 1,0^\circ)$ at $\alpha = 22^\circ$. Focusing on the controlled flows with a forcing mode at $St = 0.40$, situated near the wing tip, the effect of $C_\mu$ on the controlled tip vortex length is relatively minor, as demonstrated in figure \ref{fig:6_cmuEffects}$(a)$. Notably, an approximately $60\%$ reduction in the tip vortex length is achieved for all considered $C_\mu$ values.
	
	When examining controlled flows with a $St = 0.14$ forcing mode located inboard over the wing, an increase in $C_\mu$ results in higher time-averaged and root-mean-squared (RMS) values for $C_L$ and $C_L/C_D$, as presented in figure \ref{fig:6_cmuEffects}$(b)$. We note that a smaller $C_\mu$ could be used to reduce RMS levels and its impact on structural vibration of the wing. Given our objective to reduce the separation bubble size and enhance time-averaged lift and lift-to-drag ratio, we specifically select $C_\mu = 0.05$ as it yields superior improvements in aerodynamic performance. Importantly, even at lower $C_\mu$ values, there is a notable increase in both $\overline{C_L}$ and $\overline{C_L/C_D}$, such as a $4.73\%$ increase in $\overline{C_L}$ for $C_\mu = 0.05$ compared to $C_\mu = 0.01$. Similarly, $C_{L_\text{RMS}}$ is $0.067$ for $C\mu = 0.01$, while it increases to $0.103$ for $C_\mu = 0.05$.
	
	\subsection{A priori assessment and guidance for flow control}
	\label{sec:6_methods_resolvent_assessment}
	
	With the forcing-mode-based actuation established, let us consider an a priori assessment of the control design and its impact on the wake. To this end, we quantify mixing introduced by the modal structures using the modal streamwise, transverse and spanwise Reynolds stresses \citep{Luhar:JFM14opposition,Yeh:JFM19} respectively defined as
	\begin{equation}
	\hat{R}_x(\mathbf{x},\omega) = \text{Re}(\hat{\mathbf{q}}_{u_y}^*\hat{\mathbf{q}}_{u_z}), \quad
	\hat{R}_y(\mathbf{x},\omega) = \text{Re}(\hat{\mathbf{q}}_{u_z}^*\hat{\mathbf{q}}_{u_x}), \quad
	\hat{R}_z(\mathbf{x},\omega) = \text{Re}(\hat{\mathbf{q}}_{u_x}^*\hat{\mathbf{q}}_{u_y}) \mbox{ ,}
	\label{eq:6_modalReynoldsStress_1}
	\end{equation}
	where $\text{Re}(\cdot)$ denotes the real part of the complex-valued variable. To assess the modal capability to perturb a flow structure, we employ the following mixing metric as
	\begin{equation}
	M(\omega) = \int_{V_\text{resp}} [\sigma_1^2 (\hat{R}_x(\mathbf{x},\omega) + \hat{R}_y(\mathbf{x},\omega) + \hat{R}_z(\mathbf{x},\omega)]^{1/2} \text{d}V \mbox{ ,}
	\label{eq:6_modalReynoldsStress_2}
	\end{equation}
	where $V_\text{resp}$ denotes a specific region in space of the response modes over which the integral is computed. This region is chosen based on the flow structure targeted for modification through flow control. 
	
	\begin{figure}
		\footnotesize
		\centering
		\begin{tikzpicture}
		\node[anchor=south west,inner sep=0] (image) at (0,0) {\includegraphics[page=1,trim=4mm 0mm 4mm 0mm, clip,width=1.0\textwidth]{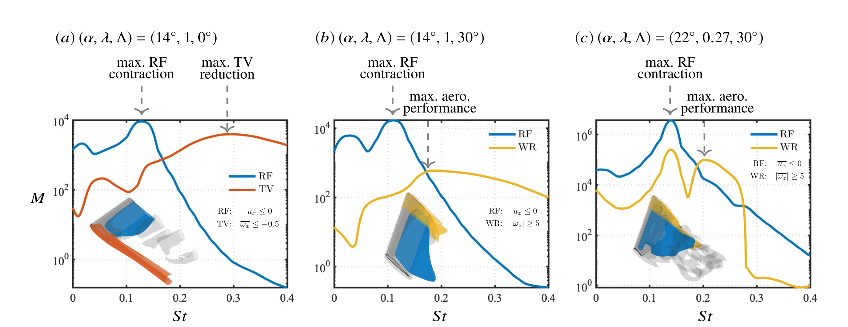}};
		\end{tikzpicture} \\ \vspace{-4mm}
		\caption{Assessment of control effects from response-mode Reynolds stress metric $M$ over $St$ for $(a)$ untapered unswept wing $(\lambda,\Lambda = 1,0^\circ)$ at $\alpha = 14^\circ$, $(b)$ untapered swept wing $(\lambda,\Lambda = 1,30^\circ)$ at $\alpha = 14^\circ$, and $(c)$ tapered swept wing $(\lambda,\Lambda = 0.27,30^\circ)$ at $\alpha = 22^\circ$. Gray arrows indicate frequencies in which controlled flows yielded maximum contraction of reversed-flow (RF) bubble, maximum tip vortex (TV) attenuation, and maximum aerodynamic performance.} 
		\label{fig:6_reynoldsStressMetric}
	\end{figure}
	
	For reducing the size of the reversed-flow bubble, let us define $V_\text{resp}$ where $\overline{u_x} \le 0$, as shown by the blue isosurfaces over both wings in figure \ref{fig:6_reynoldsStressMetric}. For the untapered unswept wing $(\lambda,\Lambda = 1,0^\circ)$, note that the peak of the blue curve in figure \ref{fig:6_reynoldsStressMetric}$(a)$ suggests that the forcing modes at $St \approx 0.14$ have higher potential to induce momentum mixing within the separation bubble, consequently achieving the control objective of reducing the reversed-flow bubble size. Conversely, when analyzing $M$ over the wing tip vortex, we define $V_\text{resp}$ to be the region for $\overline{\omega_x} \le -0.5$ and $z/c \ge 1.65$, as shown in figure \ref{fig:6_reynoldsStressMetric}$(a)$. In this case, the control objective shifts to perturbing and mitigating the tip vortex (TV, represented in red). The red curve peaks around $St \approx 0.26$, suggesting that forcing modes at this frequency can augment momentum mixing within the tip vortex, potentially weakening its core and reducing its length.
	
	Turning our attention to figure \ref{fig:6_reynoldsStressMetric}$(b)$, we focus on the untapered swept wing $(\lambda,\Lambda = 1,30^\circ)$ scenario. Here, the peak of $M$ for the blue curve suggests that a $St \approx 0.10$ forcing mode is likely to yield a larger contraction of the reversed-flow bubble. For the untapered swept wing, utilizing the tip-dominant $St = 0.10$ forcing mode substantially modifies the flow around the tip. Nevertheless, this actuation may not impact the entire reversed-flow bubble, as the upstream portion of the separated flow is located near the wing root.  To address this issue, we pursue higher frequency forcing modes that emerge near the wing root. To identify the forcing modes that enhance momentum mixing and modify the reversed-flow bubble near the wing root (WR, highlighted in yellow), we define $V_\text{resp}$ as the region where $|\overline{\omega_z}| \ge 5$ and $z/c \le 0.5$.  This span subset confines the analysis to the root region while also encompassing the upstream portion of the reversed-flow bubble. The $|\overline{\omega_z}|$ threshold only affects the magnitude of $M$ and not its distribution over $St$. We note that earlier discussions about an untapered unswept wing have excluded the WR-based analysis because it closely aligns with the blue curve for the RF-based analysis shown in figure \ref{fig:6_reynoldsStressMetric}$(a)$.
		
	For the untapered swept wing, the distribution of the WR-based $M$ suggests the utilization of a root-dominant $St = 0.18$ forcing mode for actuation. For frequencies $St \ge 0.20$, both WR-based and RF-based $M$ decrease. A rapid decline in the RF-based $M$ implies that the actuation with $St \ge 0.20$ forcing modes increases momentum mixing near the root while minimally affecting the reversed-flow bubble. Distinguishing the effects of RF- and WR-based metrics is challenging for tapered swept wings $(\lambda,\Lambda = 0.27,30^\circ)$ because the separation bubble over this wing is centered at quarter-span, as depicted in figure \ref{fig:6_reynoldsStressMetric}$(c)$. For a tapered swept wing at $\alpha = 22^\circ$, the RF-based $M$ peaks at $St = 0.14$, suggesting this mode for an actuation to reduce the reversed-flow bubble size. The WR-based metric exhibits two peaks: the first at $St = 0.14$ as the WR-based $V_\text{resp}$ encompasses part of the reversed-flow bubble. The second peak of the WR-based $M$ appears at $St = 0.20$ for a root-based actuation, while the RF-based metric $M$ remains high. This result suggests that actuation using $St = 0.20$ forcing mode can yield significant effects over the reversed-flow region.

	It is important to note that analyzing $M$ alone does not guarantee improvements in lift and lift-to-drag ratio for the controlled flows. Such enhancements need to be confirmed through experiments and numerical simulations, as demonstrated in the subsequent sections. Furthermore, we show how the behavior of controlled flows concurs with predictions made by the Reynolds stresses using the response modes. For the remainder of this manuscript, we narrow our discussion to the $3$ cases depicted in figure \ref{fig:6_reynoldsStressMetric}, as they present distinct and challenging flow features for control, including a variety of separation bubble spatial characteristics and the emergence of wing tip vortices.
	
	\subsection{Control of flow separation around untapered unswept wings}
	\label{sec:6_control_finite}
	
\begin{figure}
		\scriptsize
		\centering
		\begin{tikzpicture}
		\node[anchor=south west,inner sep=0] (image) at (0,0) {\includegraphics[page=1,trim=4mm 0mm 4mm 0mm, clip,width=1.0\textwidth]{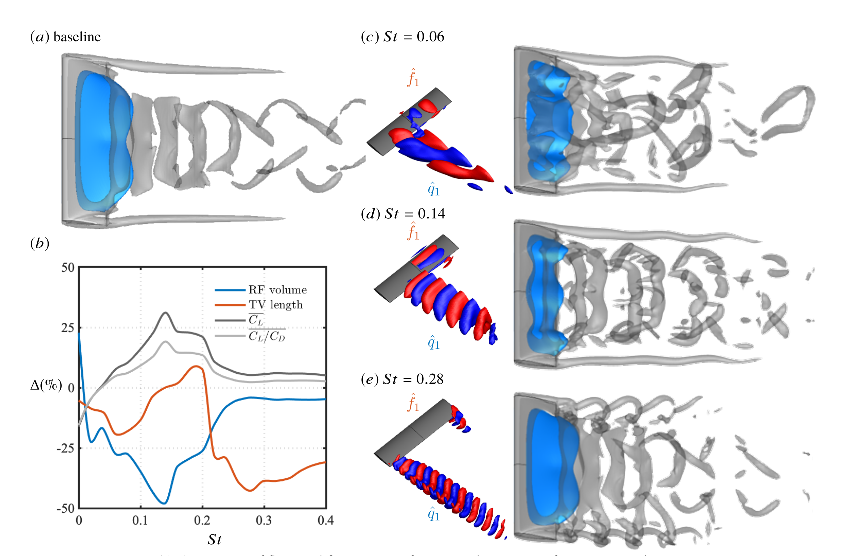}};
		\end{tikzpicture} \\ \vspace{0mm}
		\caption{Assessment of flow modification using forcing modes actuation for an untapered unswept wing $(\lambda,\Lambda = 1,0^\circ)$ at $\alpha = 14^\circ$. $(a)$ Baseline flow. $(b)$ Percentage of reversed flow (RF) volume contraction, tip vortex (TV) length reduction, and aerodynamic forces modification for controlled flows compared to baseline. $(c,d,e)$ On the left, isosurfaces of forcing $\hat{\mathbf{f}}_{uy} = \pm 1$ and response modes $\hat{\mathbf{q}}_{uy} = \pm 0.5$ at $(d)$ $St = 0.06$, $(e)$ $St = 0.14$, and $(f)$ $St = 0.28$. On the right, controlled flows. All flow fields are visualized with gray-colored isosurfaces of $Q= 1$ and blue-colored isosurfaces of $\overline{u_x} = 0$.} 
		\label{fig:6_FlowModification}
	\end{figure}

	Let us study the wake modification achieved by the present resolvent-analysis-based active flow control approach for untapered unswept wing $(\lambda,\Lambda = 1,0^\circ)$ at $\alpha = 14^\circ$. For reference, the baseline flow is visualized using gray-colored isosurfaces of $Q$, as shown in figure \ref{fig:6_FlowModification}$(a)$. Also shown over the wing is a blue isosurface representing the reversed-flow bubble, where $\overline{u_x} \le 0$. For this flow, we observe spanwise-aligned vortex shedding over the majority of the wing and a tip vortex near the free end. 
	
	The DNS of controlled flows shows substantial differences compared to baseline flows. A quantitative assessment of the control effect is shown in figure \ref{fig:6_FlowModification}$(b)$. Here, the percentage difference ($\Delta$) is used to illustrate variations in flow features between the baseline and controlled flows. In particular, let us analyze the blue curve that represents the contraction (negative) and expansion (positive) effect of flow control on the reversed-flow bubble, which is one of the objectives of the present study. The minimum reversed-flow bubble volume is achieved when forcing mode at $St = 0.14$ is used for actuation. At this frequency, the reversed-flow bubble undergoes a $50\%$ volume reduction, as seen in figure \ref{fig:6_FlowModification}$(d)$.
	
	The notable reduction in the reversed-flow bubble size at $St = 0.14$ leverages the forcing-response mode pair at this frequency. In particular, the response mode emerges over the baseline separation bubble, indicating that the forcing-mode actuation excites structures within the reversed-flow bubble. Indeed, along the spanwise-aligned root vortices, the controlled flow using forcing modes at $St = 0.14$ induces spatial oscillations in the wake that are absent in the baseline flow, as seen in \ref{fig:6_FlowModification}$(d)$. These structures perturb the baseline reversed-flow bubble and result in its volumetric contraction.
	
	Concurrently, the controlled flow with $St = 0.14$ forcing-mode actuation exhibits the highest increase in lift and lift-to-drag ratio, shown by the percentage difference in $\overline{C_L}$ and $\overline{C_L/C_D}$ between the baseline and controlled flows (gray curves). We note that improvements in the aerodynamic performance are achieved because the amount of lift increase is substantially greater than that of the drag. The concomitant $\overline{C_L/C_D}$ increase and reversed-flow bubble contraction shown in figure \ref{fig:6_FlowModification}$(b)$ indicate that diminishing the reversed-flow size enhances the overall aerodynamic performance of the wing. As the actuation frequency increases, the higher spatial support of the forcing-mode pairs shifts toward the wing tip and the modes emerges outside the reversed-flow region. As a result, the effect of contracting the reversed-flow bubble is minimized for high-frequency actuation, reaching a plateau for $St \ge 0.26$.
	
	Forcing-mode actuation for $St \ge 0.26$ is also important for flows over untapered unswept wings $(\lambda,\Lambda = 1,0^\circ)$ due to their ability to control of the tip vortex while moderately increasing lift and lift-to-drag ratio, as seen in figure \ref{fig:6_FlowModification}$(b)$. The attenuation and control of tip vortices has been extensively studied \citep{Gursul:AFC07tipcontrol,Greenblatt:AIAAJ12,Gursul:AIAAJ18} due to its adverse effects on wing aerodynamics \citep{Francis:JA79,Katz:JA1989roughness,Green:JFM91,Devenport:JFM96} and its negative impact in air transportation \citep{Spalart:ARFM98}. In prior studies involving laminar steady flow around an untapered unswept wing, a $21\%$ reduction was achieved using instability analysis modes as body forces \citep{Edstrand:JFM18a,Edstrand:JFM18b}. For this approach, a vortex instability was taken advantage of to increase dissipation of the tip vortex.
	
	While the $St = 0.14$ forcing mode proves highly effective in diminishing the volume of the separation bubble, its impact on the reduction of the tip vortex length is relatively modest, as seen in the red curve of figure \ref{fig:6_FlowModification}$(b)$. Here, the control effect on the tip vortex is measured through the percentage difference between the baseline and the controlled tip vortex lengths. For figure \ref{fig:6_FlowModification}$(b)$, the tip vortex length is estimated using the distance in the streamwise direction between the trailing edge and the tip of the $\overline{\omega_x} = -1$ isosurface. While the chosen value of the $\overline{\omega_x}$ isosurface is arbitrary, it does not affect the findings. 
	
	A decrease in the tip vortex length is achieved for the control using $St \le 0.10$ forcing modes, reaching a local minimum at $St = 0.06$. The forcing-response mode pair at $St = 0.06$ and the corresponding controlled flow are depicted in figure \ref{fig:6_FlowModification}$(c)$. This actuation induces prominent streamwise vortices over the inboard region, perturbing the reversed-flow bubble and diminishing its volume. These structures also interact with the downstream portion of the tip vortex, resulting in an overall reduction in its length.
	
	To achieve a more substantial weakening of the tip vortex, forcing modes at higher frequencies are employed for direct wake modification. The controlled flows with $St \ge 0.26$ forcing modes achieve a higher level of reduction in the tip vortex length, while preserving the baseline reversed-flow bubble, as depicted in figure \ref{fig:6_FlowModification}$(e)$ for the controlled flow with $St = 0.28$ forcing mode Near the wing tip, structures excited by the forcing mode emerge predominantly from the pressure side of the wing tip and induce a helical vortical structure formation that is associated with a significant reduction of the tip vortex length. Moreover, the weakening of the tip vortex alleviates the inboard downwash, resulting in a $6\%$ overall lift increase.
	
\begin{figure}
		\footnotesize
		\centering
		\begin{tikzpicture}
		\node[anchor=south west,inner sep=0] (image) at (0,0) {\includegraphics[page=1,trim=4mm 2mm 6mm 0mm, clip,width=1.0\textwidth]{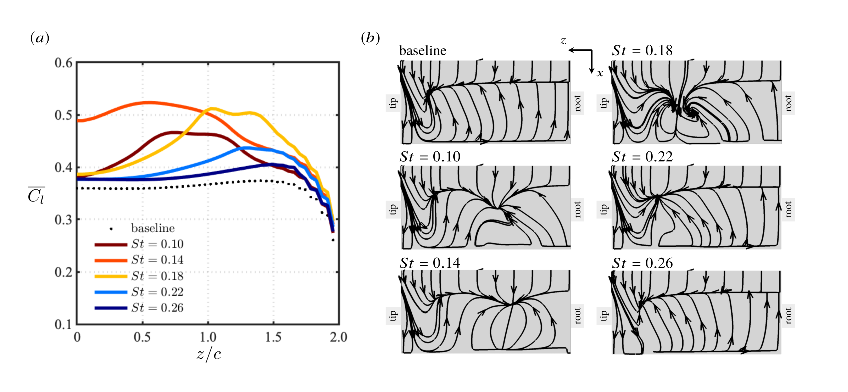}};
		\end{tikzpicture} \\ \vspace{-4mm}
		\caption{$(a)$ Sectional lift $C_l$ distribution over span for an untapered unswept wing $(\lambda,\Lambda = 1,0^\circ)$ at $\alpha = 14^\circ$. $(b)$ Time-averaged skin friction lines over the suction side of the wing.} 
		\label{fig:6_liftControl}
	\end{figure}

	Figure \ref{fig:6_liftControl}$(a)$ shows the sectional lift ($C_l$) over the wingspan for baseline and controlled flows at different actuation frequencies. The $z/c$ region where $C_l$ increases, compared to the baseline flow, coincides with the spanwise region where the forcing-response mode pairs have higher spatial support. For instance, for the $St = 0.10$ forcing mode actuation, the most pronounced increase in $C_l$ is observed between $0.5 \le z/c \le 1.2$, contributing to a $16\%$ overall lift increase.  Similarly, for the controlled flow with $St = 0.14$ forcing mode, a larger rise in $C_l$ is evident over a broader portion of the wingspan between $0 \le z/c \le 1.5$, resulting in $31\%$ total lift increase. For the controlled flow with $St = 0.18$ forcing mode, the greatest increase in lift occurs over a narrower wingspan region compared to the control with $St = 0.14$. For the control with $St = 0.18$ forcing mode, lift increases especially closer to the wing tip between $0.7 \le z/c \le 1.7$, resulting in a $22\%$ increase in the overall lift.
		
\begin{figure}
		\footnotesize
		\centering
		\begin{tikzpicture}
		\node[anchor=south west,inner sep=0] (image) at (0,0) {\includegraphics[page=1,trim=4mm 0mm 9mm 0mm, clip,width=1.0\textwidth]{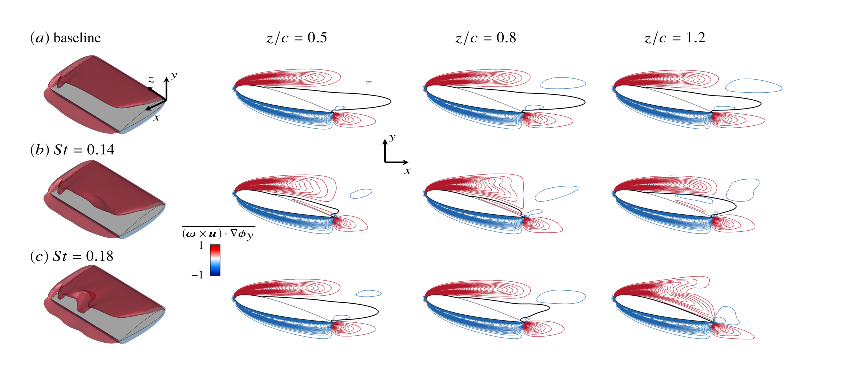}};
		\end{tikzpicture} \\ \vspace{-4mm}
		\caption{Lift elements around an untapered unswept wing $(\lambda,\Lambda = 1,0^\circ)$ at $\alpha = 14^\circ$. $(a)$ baseline, $(b)$ controlled flows with  $St = 0.14$ and $(c)$ $St = 0.18$ forcing modes. Leftmost figures show $3$-D time-averaged lift elements with isosurfaces of $\overline{(\bs{\omega} \times \mathbf{u}) \cdot \nabla \phi_y} = \pm 1$. To the right of the $3$-D view, slices at $z/c = 0.5$, $0.8$, and $1.2$ show the red and blue contours of lift elements and a black line contour of $\overline{u_x} = 0$.} 
		\label{fig:6_liftForceElements}
	\end{figure}

	The $C_l$ increase results from vortical lift from near-wake structures excited by the actuation. The signature of these vortical structures advecting over the wing marks the skin friction field over their suction side, as seen in figure \ref{fig:6_liftControl}$(b)$, showing that separation is not suppressed. Using force element analysis (details in appendix \ref{sec:6_appendix_forceelements}), we reveal how such near-wake structures contribute to lift in figure \ref{fig:6_liftForceElements}. Drag elements distribution, qualitatively similar to lift elements, are not shown for brevity. Lift elements show that the major lift contribution (positive, red) comes from the flow structures located outside of the reversed-flow region over a significant portion of the wingspan.
		
	For force element analysis, the volume force elements  are identified by the dot product of the Lamb vector ($\bs{\omega} \times \mathbf{u}$) and the gradient of the auxiliary potential ($\nabla \phi_i$). The auxiliary potential field decays rapidly in magnitude far from the wing surface. The $\nabla \phi_i$ characteristics are only affected by the wing planform, being consistent for baseline and controlled flows. For this reason, a strategic  wake modification that reduces the reversed-flow bubble size has the potential to bring vortices closer to the wing where $\nabla \phi_i$ has a higher magnitude.  Consequently, flow structures emerging in controlled flows near the wing have higher contribution to the overall vortical~lift. It is important to note that the higher magnitude of $\nabla \phi_i$ consistently appears near the wing surface for all wing planforms and local lift enhancements result from the strengthening of lift elements caused by the flow field. For instance, for flows over untapered and tapered swept wings, \cite{Zhang:PRF22} and \cite{Ribeiro:JFM2023tapered} have shown that major increases in root contribution to the overall lift results from the emergence of root-based vortices.
	
	As shown in figures \ref{fig:6_liftForceElements}$(a,b)$ for the baseline and controlled flow with the $St = 0.14$ forcing mode, the reversed-flow bubble size significantly decreases at $z/c = 0.5$ and $0.8$. This causes lift elements to emerge nearer to the wing compared to the baseline. Consequently, the highest lift increase occurs near the inboard quarter-span of the wing, as depicted by the red curve in figure \ref{fig:6_liftControl}$(a)$. For the controlled flow with $St = 0.18$, the reversed-flow bubble reduces significantly over the outboard portion of the wingspan at $z/c = 1.2$ when compared to the baseline case, as depicted in figures \ref{fig:6_liftForceElements}$(a,c)$. This reversed-flow structure contraction at $z/c = 1.2$ causes lift elements to emerge over this region and closer to the wing. As a result, this flow modification leads to a substantial lift increase over the outboard quarter-span of the wing, as seen in the yellow curve in figure \ref{fig:6_liftControl}$(a)$. The control effects resulting from actuation at the shedding frequency over an untapered unswept wing $(\lambda,\Lambda = 1,0^\circ)$ at $\alpha = 14^\circ$ are analogous to the ones observed at $\alpha = 22^\circ$ (discussions are omitted for brevity).
	
	\subsection{Control of flow separation around untapered swept wings}
	\label{sec:6_control_swept}	
\begin{figure}
		\scriptsize
		\centering
		\begin{tikzpicture}
		\node[anchor=south west,inner sep=0] (image) at (0,0) {\includegraphics[page=1,trim=4mm 0mm 1mm 0mm, clip,width=1.0\textwidth]{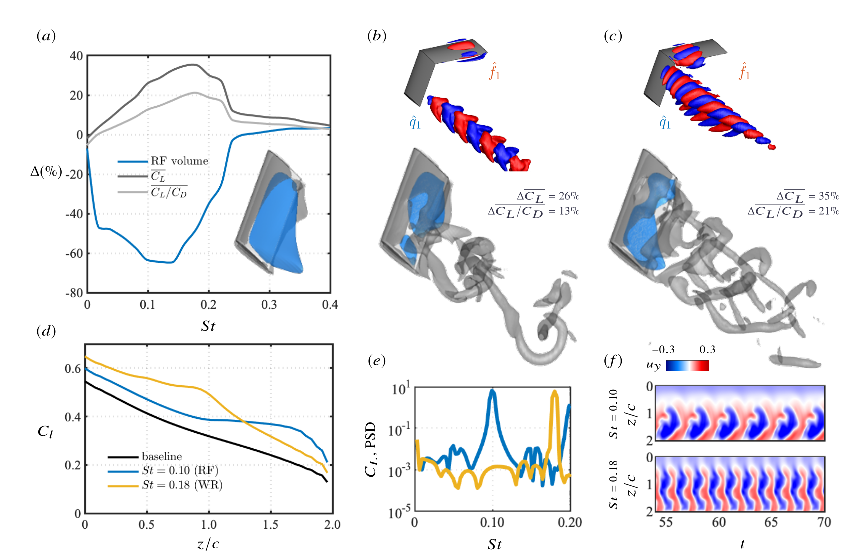}};
		\end{tikzpicture} \\ \vspace{-4mm}
		\caption{Assessment of flow modification using forcing modes actuation for an untapered swept wing $(\lambda,\Lambda = 1,30^\circ)$. Control with forcing modes located over the reversed-flow (RF) bubble and wing root (WR) are covered in detail. $(a)$ RF volume contraction and aerodynamic changes for controlled flows. Baseline flow on the bottom-right corner. $(b,c)$ On top, isosurfaces of forcing $\hat{\mathbf{f}}_{uy} = \pm 1$ and response modes $\hat{\mathbf{q}}_{uy} = \pm 0.5$ at $(b)$ $St = 0.10$ and $(c)$ $St = 0.18$. On the bottom, instantaneous controlled flows over half span. All flows are visualized with gray-colored isosurfaces of $Q= 1$ and blue-colored isosurfaces of $\overline{u_x} = 0$. $(d)$ $C_l$ over wingspan. $(e)$ lift PSD and $(f)$ probed $u_y$ for controlled flows.} 
		\label{fig:6_sweptControl}
	\end{figure}

	For  separated flows over untapered swept wings $(\lambda,\Lambda = 1,30^\circ)$, our objective is to directly modify the wake and control the size of the reversed-flow bubble. Unlike the findings presented in Section \ref{sec:6_control_finite}, achieving the maximum reduction in the volume of the reversed-flow bubble around swept wings may not necessarily lead to optimal lift and lift-to-drag ratio enhancements. To illustrate the control effects, let us examine the laminar separated flow around a swept wing with $\Lambda = 30^\circ$ at $\alpha = 14^\circ$. The steady baseline flow is depicted in the insert visualization of figure \ref{fig:6_sweptControl}$(a)$. The reversed-flow region is highlighted with a blue isosurface at $\overline{u_x} = 0$. The application of resolvent-analysis-based actuation results in unsteady wakes that modify the $3$-D dynamics about the reversed-flow bubble, contracting its volume.  
	
	The baseline reversed-flow bubble is larger near the tip. Consequently, to achieve the most significant reduction in the reversed-flow bubble volume, direct wake modification is applied near the wing tip. In this regard, the flow field is actuated using a forcing mode in the frequency range of $0.10 \le St \le 0.14$, whose spatial support is predominantly concentrated near the wing tip. As shown in figure \ref{fig:6_sweptControl}$(b)$, the controlled flow at these frequencies induces a local shedding close to the tip, exciting structures that perturb and modify the reversed-flow bubble. Given the actuation near the wing tip, the inboard flow near the wing root remains undisturbed. 
	
	The reduction of the reversed-flow bubble size, through actuation with $0.10 \le St \le 0.14$ forcing mode, results in a significant improvement in lift and lift-to-drag ratio. The most pronounced aerodynamic enhancements in $\overline{C_L}$ and $\overline{C_L/C_D}$ are achieved when promoting structures near the wing root, where the baseline flow is attached to the wing surface. To actuate in this region,  we employ the forcing mode at $St = 0.18$, which exhibits larger spatial support near the wing root, as depicted in figure \ref{fig:6_sweptControl}$(c)$.  This actuation substantially increases the root contribution to the overall lift, as seen in figure \ref{fig:6_sweptControl}$(d)$. As discussed for controlled flows over the untapered unswept wing $(\lambda,\Lambda = 1,0^\circ)$ in section \ref{sec:6_control_finite}, the strengthening of near-surface vortices over a section of the wingspan yields sectional lift enhancements. The coherent structures promoted by the actuation mechanism produce the $C_l$ increase near the tip for the controlled flow with the tip-based forcing mode at $St = 0.10$. In contrast, the actuation with the root-based forcing mode at $St = 0.18$ produces higher increase in $C_l$ near the root.
	
\begin{figure}
		\scriptsize
		\centering
		\begin{tikzpicture}
		\node[anchor=south west,inner sep=0] (image) at (0,0) {\includegraphics[page=1,trim=4mm 3mm 11mm 0mm, clip,width=1.0\textwidth]{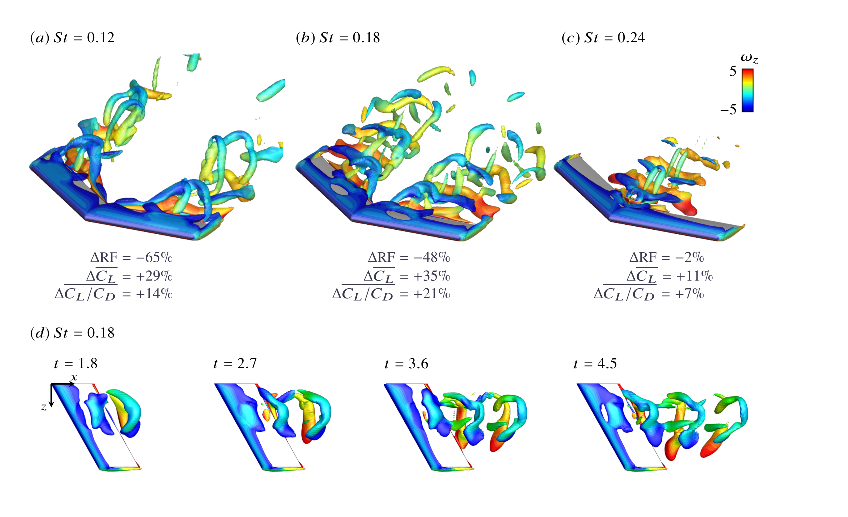}};
		\end{tikzpicture} \\ \vspace{-4mm}
		\caption{Instantaneous visualizations of flow modification for controlled flows with $(a)$ $St = 0.12$, $(b)$ $St = 0.18$, and $(c)$ $St = 0.24$ forcing mode actuation for untapered swept wing with $\Lambda = 30^\circ$ at $\alpha = 14^\circ$. Inserts of percentual difference ($\Delta$) of time-averaged reversed-flow volume (RF), $\overline{C_L}$, and $\overline{C_L/C_D}$ with respect to baseline flow. $(d)$ shows the inital evolution of the unsteady controlled flow at $St = 0.18$, for $1.8 \le t \le 4.5$. Flow structures are visualized with isosurfaces of $Q = 2$ colored with ${\omega_z}$.} 
		\label{fig:6_sweptControl_2}
	\end{figure}

	The lift and lift-to-drag ratio enhancements achieved through control using the $St = 0.18$ forcing mode surpass those obtained with other actuation frequencies. In fact, controlled flows with actuation frequencies $St \ge 0.20$ experience a significant decline in lift and lift-to-drag ratio improvements, despite utilizing a spatial support over the wing root similar to that of the $St=0.18$ forcing mode. Moreover, as depicted in the blue curve of figure \ref{fig:6_sweptControl}$(a)$, controlled flows using $St \ge 0.24$ forcing modes increase the size of the reversed-flow bubble. As illustrated in figure \ref{fig:6_sweptControl_2}($a$-$c$), which compares controlled flows using $St=0.12$, $0.18$, and $0.24$, only actuation using $St = 0.18$ forcing mode induces a global wake modification that alters flow dynamics across the entire wingspan. Both lower- and higher-frequency forcing modes primarily alter flow in specific regions where actuation occurs, near the wing tip for $St=0.12$ and near the root for $St=0.24$. Controlled flows with $St \ge 0.24$ forcing modes yield minor changes on the reversed flow because their spatial support is concentrated near the wing root region and do not merge into the separated flow bubble. The aerodynamic modifications achieved with $St \ge 0.24$ are not related to any changes in the reversed flow bubble but are mainly due to the emergence of near-surface vortices excited by the actuation mechanism. As shown in figure \ref{fig:6_sweptControl_2}$(d)$, for the controlled flow with $St =0.18$, the global wake modification is initiated by the perturbation, arising near the wing root and rapidly evolving downstream into hairpin-like vortices that reshape the overall wake dynamics. Consequently, the controlled flow using the $St=0.18$ forcing mode benefits from a local lift increase around the root-based actuation and a subsequent reduction in the reversed-flow bubble, leading to significant aerodynamic improvements.
	
	The periodic controlled flows synchronizes with the actuation frequency, as shown by the PSD of $C_L$ in figure \ref{fig:6_sweptControl}$(e)$. Here, the lift spectra peaks at the input frequencies and its harmonics for both $St=0.10$ (blue) and $St = 0.18$ (yellow) controlled flows. The actuation frequency affects the size of flow structures, as the higher-frequency input yields elongated spanwise-aligned vortices, as shown by the probed $u_y$ along $(x,y)/c = (3,-0.5)$ over time in figure \ref{fig:6_sweptControl}$(f)$. Additionally, the higher-frequency actuation yields a lower lift-to-drag RMS. While $(C_L/C_D)_\text{RMS} = 0.148$, for the controlled flow at $St = 0.10$, the root-based controlled flow at $St = 0.18$ results in $(C_L/C_D)_\text{RMS} = 0.079$. For brevity, only the control effects for the untapered swept wing $(\lambda,\Lambda = 1,30^\circ)$ at $\alpha = 14^\circ$ were presented here. We note that the control effects are analogous for both angles of attack. At $\alpha = 22^\circ$, the control using a lower-frequency mode near the wing tip is also less effective in improving lift and aerodynamic performance than a higher-frequency forcing mode located near the wing root (see figure \ref{fig:6_schematics}).

	\subsection{Control of flow separation around tapered swept wings}
	\label{sec:6_control_taper}	
	For post-stall flows over tapered swept wings, our aim is to control the size of the reversed-flow bubble. Flow control effects over tapered swept wings are consistent with those observed for untapered swept wings, as discussed in section \ref{sec:6_control_swept}. For a tapered swept wing $(\lambda,\Lambda = 0.27,30^\circ)$ at $\alpha = 22^\circ$, the actuation frequency that maximally reduces the reversed-flow bubble size does not coincide with the frequency leading to the highest aerodynamic improvements  in $\overline{C_L}$ and $\overline{C_L/C_D}$.
	
\begin{figure}
		\scriptsize
		\centering
		\begin{tikzpicture}
		\node[anchor=south west,inner sep=0] (image) at (0,0) {\includegraphics[page=1,trim=4mm 0mm 4mm 0mm, clip,width=1.0\textwidth]{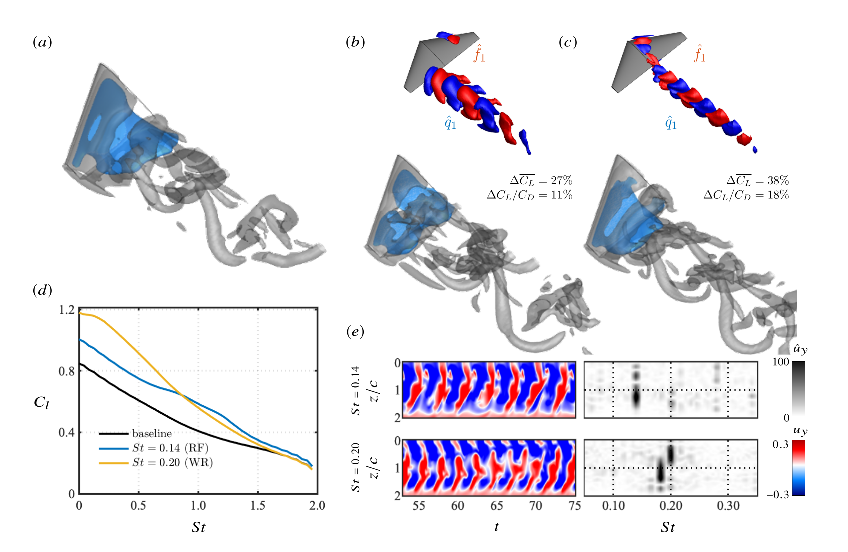}};
		\end{tikzpicture} \\ \vspace{-4mm}
		\caption{Assessment of flow modification using forcing modes actuation for tapered swept wing $(\lambda,\Lambda = 0.27,30^\circ)$ at $\alpha = 22^\circ$. $(a)$ Baseline flow. $(b,c)$ On top, isosurfaces of forcing $\hat{\mathbf{f}}_{uy} = \pm 1$ and response modes $\hat{\mathbf{q}}_{uy} = \pm 0.5$ at $(b)$ $St = 0.10$ and $(c)$ $St = 0.18$. On the bottom, instantaneous controlled flows over half span. All flows are visualized with gray-colored isosurfaces of $Q= 1$ and blue-colored isosurfaces of $\overline{u_x} = 0$. Improvements in $\overline{C_L}$ and $\overline{C_L/C_D}$ are shown. $(d)$ $C_l$ over wingspan. $(e)$ Temporal behavior of probed $u_y$ for controlled flows.} 
		\label{fig:6_taperControl_b30a22s30}
	\end{figure}

	The baseline flow exhibits high levels of wake fluctuations across a significant portion of the wingspan, as shown in figure \ref{fig:6_taperControl_b30a22s30}$(a)$.  The spatial support of the resolvent modes employed for flow control is also distributed over the wingspan. Forcing-response mode pairs are primarily concentrated over the quarter-span ($St \le 0.18$), the wing root ($0.18 < St < 0.28$), and the wing tip ($St \ge 0.28$). Actuation using forcing modes at frequencies $St \ge 0.28$ introduces flow perturbations near the tip region but proves ineffective in modifying the wake or enhancing the overall aerodynamic performance. This is attributed to the absence of a noticeable tip vortex structure in the baseline flow.
	
	For the tapered swept wing $(\lambda,\Lambda = 0.27,30^\circ)$, based on the characteristics of the modal structures and insights gained from flow control discussions in sections \ref{sec:6_control_finite} and \ref{sec:6_control_swept}, we assess the control effects at only two frequencies, $St = 0.14$ and $0.20$. We recall that the largest $(x,y)$ sectional area of the baseline separation bubble for this wing is located at the quarter-span (see section \ref{sec:6_baseline_timeaveraged}). For the first actuation frequency using the $St = 0.14$ forcing mode,  whose spatial support is present near the quarter-span and within the reversed-flow bubble, perturbations are excited within the reversed-flow bubble, modifying the near-wake and reducing the reversed-flow volume, as seen in figure \ref{fig:6_taperControl_b30a22s30}$(b)$. 
	
	The second frequency of actuation involves wake modification near the wing root region, aiming to assess whether a more significant enhancement  in $\overline{C_L}$ and $\overline{C_L/C_D}$ can be achieved by perturbing this area, akin to the effects observed for swept wings. The forcing mode at $St = 0.20$ exhibits higher spatial support over the root region, as seen in figure \ref{fig:6_taperControl_b30a22s30}$(c)$. The DNS of the controlled flows at this frequency promotes structures near the wing root, leading to an increased root contribution to the overall lift, as depicted in figure \ref{fig:6_taperControl_b30a22s30}$(d)$. Additionally, this actuation results in a greater increase in lift and lift-to-drag ratio compared to the lower-frequency actuation at $St = 0.14$.
	
	The differences in the wake patterns between baseline and controlled flows are not readily apparent from the $3$-D depiction of coherent structures in figures  \ref{fig:6_taperControl_b30a22s30}$(b,c)$. Additionally, controlled flows with $St = 0.14$ and $0.20$ forcing modes achieve approximately $38 \%$ of contraction of the reversed-flow volume. To delve into the distinctions between baseline and controlled flows, we examine the probed $u_y$ at $(x,y)/c = (3,-0.5)$ over time and frequency in figure \ref{fig:6_taperControl_b30a22s30}$(e)$. While omitted here for brevity, we note that the baseline and actuated wakes using the forcing mode at $St = 0.14$ exhibit minimal differences in both temporal and spectral analyses of the probed data.
	
	Notable differences emerge when comparing the probed data of controlled flows at $St = 0.14$ and $0.20$ as the peak frequencies of the controlled flow with the $St = 0.20$ forcing mode shift to a higher $St$ range. Wake vortices exhibit a spanwise split into two shedding modes: near-root vortices synchronized with the actuation frequency and quarter-span vortices with a lower frequency unsynchronized with the actuation frequency. This behavior induces a low-frequency beating in the lift spectrum. This phenomenon, known as vortex dislocation, has been observed in the wakes of high-aspect-ratio blunt bodies and controlled flows \citep{Eisenlohr:PF89vortexDislocation,Williamson:JFM89oblique,Zhang:JFM20}. Despite the lower aspect ratio in the present controlled flow, a similar wake pattern emerges due to the interplay between near-tip and near-root shedding structures. Lastly, we note that both $St = 0.14$ and $St = 0.20$ forcing-mode actuation strategies increase lift-to-drag RMS, which is originally $(C_L/C_D)_\text{RMS} = 0.017$ for the baseline flow. For the controlled flow with the $St = 0.14$ forcing mode, $(C_L/C_D)_\text{RMS}$ increases to $0.098$, while the root-based actuation at $St = 0.20$ yields a lower $(C_L/C_D)_\text{RMS} = 0.059$. Again, we note that the control effects for the tapered swept wing at $\alpha = 14^\circ$ and $22^\circ$ are analogous and we have omitted the results for the lower incidence angle for brevity.
	
	\subsection{Control of tip vortex around untapered unswept wings}
	\label{sec:6_control_tip}
	To understand the mechanisms that lead to the tip vortex attenuation, we focus our discussion on the flow over an untapered unswept wing $(\lambda,\Lambda = 1,0^\circ)$ at $\alpha = 14^\circ$. We compute the streamwise circulation $\Gamma_x = \int_{C} = \mathbf{u} \cdot \text{d}\mathbf{l}$, where $C$ is the isocontour of $\overline{\omega_x} = -1$ on the $(y,z)$ plane at varied $x$. To compute $\Gamma_x$, only the isocontours of $\overline{\omega_x} = -1$ aligned at the wing tip at $z/c \approx 2$ are considered. In this manner, we limit the inboard vorticity influence on $\Gamma_x$, caused by the interplay of the tip vortex and the root shedding structures. Our discussions focus on wake modifications relative to the baseline and are not significantly influenced by the specific value of the $\overline{\omega_x}$ isocontour used to compute the circulation.
		
\begin{figure}
		\footnotesize
		\centering
		\begin{tikzpicture}
		\node[anchor=south west,inner sep=0] (image) at (0,0) {\includegraphics[page=1,trim=4mm 2mm 10mm 0mm, clip,width=1.0\textwidth]{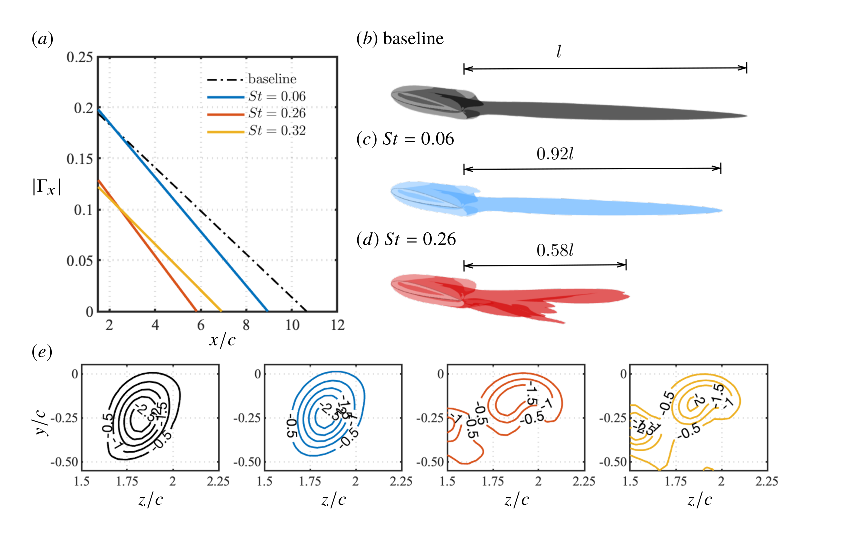}};
		\end{tikzpicture} \\ \vspace{-4mm}
		\caption{Tip vortex attenuation for optimal forcing modes actuation at specific frequencies for untapered unswept wing $(\lambda,\Lambda = 1,0^\circ)$ at $\alpha = 14^\circ$. Throughout the figure, color black is used for baseline case, blue is used for $St = 0.06$, red is used for $St = 0.26$, and yellow for $St = 0.32$ forcing mode actuation. $(a)$ Tip vortex streamwise circulation $|\Gamma_x|$ over $x/c$. $(b,c,d)$ Side view of the time-averaged flow structures using isosurfaces of $\overline{Q} = 1$. $(e)$ Streamwise vorticity contours shown in $2$-D slices ($(y,z)$ plane) at $x/c = 3$. } 
		\label{fig:6_tipControl}
	\end{figure}

	Let us analyze the specifics of the $St = 0.06$ forcing mode actuation. This actuation results in $8\%$ reduction in tip vortex length, as shown in figure \ref{fig:6_tipControl}$(c)$, compared to the baseline length shown in \ref{fig:6_tipControl}$(b)$. As seen in the vorticity contours at $x/c = 3$ in figure \ref{fig:6_tipControl}$(e)$, this low-frequency actuation has minimal impact on the core of the tip vortex near the wing because its perturbations arise over the inboard region of the flow, increasing unsteadiness locally. This inboard wake interacts with the tip vortex downstream increasing its dissipation, as shown by the higher decay rate of circulation $\mathrm{d}|\Gamma_x|/\mathrm{d}x$, resulting in an overall reduction in the tip vortex length.
	
	When employing tip-based actuation at $St \ge 0.22$, the resulting flows exhibit a substantial attenuation of the tip vortex. The controlled flow with the $St = 0.26$ forcing mode achieves a remarkable reduction in tip vortex length, reaching approximately $42\%$, as depicted in figure \ref{fig:6_tipControl}$(d)$. Examining the  vorticity contours at $x/c=3$ in figure \ref{fig:6_tipControl}$(e)$, we reveal that the actuation associated with the $St=0.26$ forcing mode effectively modifies the vortex core near the wing. This modification is predominantly observed in the lower portion of the vortex core between $-0.5 \le y/c \le -0.25$, as a consequence of this forcing mode actuation being localized on the pressure side of the wing.
	
	For controlled flows with $St = 0.32$, circulation near the wing is reduced, as illustrated by the vorticity contours at $x/c=3$. However, this forcing mode actuation is suboptimal for reducing the tip vortex length downstream in the wake. The structures excited by the $St = 0.32$ actuation remain coherent downstream, resulting in a lower circulation decay rate. It is noteworthy that the resolvent-analysis-based control achieves substantial tip vortex attenuation for $St \ge 0.26$ while also maintaining the increases in $\overline{C_L}$, $\overline{C_D}$, and $\overline{C_L/C_D}$ within $6\%$ difference from the baseline flow, as demonstrated in table \ref{tab:6_liftAssessment}.
	
	\renewcommand{\arraystretch}{1.2}
	\begin{table}
		\vspace*{0mm}
		\centering
		\begin{tabular}{p{1.15in}p{0.01in}p{0.48in}p{0.01in}p{0.48in}p{0.01in}p{0.48in}p{0.01in}p{0.48in}p{0.01in}p{0.48in}p{0.01in}p{0.48in}}
			Cases && 
			\multicolumn{1}{l}{\hspace{0mm}$\overline{C_L}$} && \multicolumn{1}{l}{\hspace{0mm}$C_{L_\text{RMS}}$} && 
			\multicolumn{1}{l}{\hspace{0mm}$\overline{C_D}$} && \multicolumn{1}{l}{\hspace{0mm}$C_{D_\text{RMS}}$} &&
			\multicolumn{1}{l}{\hspace{0mm}$\overline{C_L/C_D}$} && \multicolumn{1}{l}{\hspace{0mm}$(C_L/C_D)_{_\text{RMS}}$} \\ 
			Baseline && $0.713$ && $0.016$ && $0.447$ && $0.002$ && $1.594$ && $0.027$  \\
			Controlled ($St = 0.26$) && $0.755$ && $0.037$ && $0.461$ && $0.009$ && $1.637$ && $0.052$ \\
			Controlled ($St = 0.30$) && $0.753$ && $0.034$ && $0.460$ && $0.008$ && $1.634$ && $0.048$ \\
			Controlled ($St = 0.34$) && $0.755$ && $0.028$ && $0.460$ && $0.006$ && $1.640$ && $0.042$ \\
			Controlled ($St = 0.38$) && $0.753$ && $0.025$ && $0.459$ && $0.005$ && $1.640$ && $0.039$ \\
		\end{tabular} \vspace*{0mm}
		\caption{Effect of tip vortex control using $St \ge 0.26$ forcing modes on time-averaged and RMS of $C_L$, $C_D$, and $C_L/C_D$.}
		\label{tab:6_liftAssessment}
	\end{table}	
	
	Similar results were obtained when investigating tip vortex attenuation for an untapered unswept wing $(\lambda,\Lambda = 1,0^\circ)$ at $\alpha = 22^\circ$. For this case, the optimal actuation for tip vortex attenuation is identified for the tip-dominant forcing mode at $St = 0.40$, leading to an sizeable reduction in the tip chord length of approximately $60\%$. In the Reynolds number regime considered herein, the analysis of tip vortex attenuation is conducted for untapered unswept wings, given that tapered swept wings $(\lambda,\Lambda = 0.27,30^\circ)$ lack a distinct streamwise vortex core at the wing tip, as shown in figure \ref{fig:6_dns}.
	
	\section{Conclusions}
	\label{sec:6_conclusions}
	We performed direct numerical simulations involving external actuation with optimal forcing modes identified through triglobal resolvent analysis. The baseline flows for all cases exhibited a large stalled flow region on the suction side of the wing. Our investigation identifies structures capable of modifying the separated wake, consequently enhancing the overall lift of the wing. For laminar separated flows, the emergence of coherent structures near the wing was found to positively impact aerodynamics. Active control through direct wake modification was demonstrated to improve the lift and lift-to-drag ratio of laminar post-stall flows over low-aspect-ratio wings, with the specific strategy depending on the wing planform and the reversed-flow bubble topology.
	
	For an untapered unswept wing $(\lambda,\Lambda = 1,0^\circ)$, actuation near the root region with frequency content close to that of the baseline vortex shedding, significantly reduced the reversed-flow volume. With the bubble contraction, this actuation induced unsteady vortical structures closer to the wing, leading to a substantial increase in overall lift. At higher frequencies, optimal forcing modes appeared in the vicinity of the wing tip, originating from the trailing edge at the pressure side. Flows actuated using these modes exhibited an unsteady helical vortex formation, suppressing the quasi-steady streamwise vortex core seen in the baseline flow. While the tip vortex circulation decayed at any distance from the wing, modest effects were observed on the inboard wake.
	
	The application of direct wake modification to untapered and tapered swept wings $(\Lambda = 30^\circ)$ was also guided by the insights of forcing-response mode pairs and baseline flow physics. High-frequency actuation using forcing modes with higher spatial support near the wing root of both untapered and tapered swept wings induced unsteadiness that modified the near wake, leading to increased lift and lift-to-drag ratio. Our study also highlighted the utility of response modes and their modal Reynolds stresses for an a priori assessment of control effects. Overall, our findings underscore the receptiveness of laminar separated flows to external perturbations, demonstrating that the optimal spatial-temporal input can significantly alter the $3$-D dynamics of the wake and improve the aerodynamic performance of the wing.
	
	\appendix
	\section{Force element analysis}
	\label{sec:6_appendix_forceelements}
	Force element analysis \citep{Chang:PRSA92} identifies flow structures that exert aerodynamic loads on the wing. This method shares close similarities with force elements derived through a variational approach \citep{Quartapelle:AIAAJ83}, force partition methods \citep{Menon:JCP21}, and many others. In the present approach, we define an auxiliary potential with boundary condition of $-\mathbf{n} \cdot \nabla \phi_i = \mathbf{n} \cdot \mathbf{e}_i$ set on the wing surface, where $\phi$ is the auxiliary potential, $\mathbf{n}$ is the unit wall normal vector, and $\mathbf{e}_i$ is the unit vector in the $i$th direction. By taking the inner product of the Navier--Stokes equations with $\nabla \phi$ and performing an integral over the fluid domain, the forces exerted in the $i$-th direction can be expressed as
	\begin{equation}
	F_i = \int_V \bs{\omega} \times \mathbf{u} \cdot \nabla \phi_i \text{d}V + \frac{1}{Re} \int_S \bs{\omega} \times \mathbf{n} \cdot (\nabla \phi_i + \mathbf{e_i}) \text{d}S,
	\label{eq:6_forceelements}
	\end{equation}
	where the first integrand represents the volume force elements and the second integrand corresponds to the surface force elements. To illustrate the lift elements in the flow field, we take the Hadamard product of the gradient of the auxiliary potential $\nabla \phi_i$ and the Lamb vector ($\bs{\omega} \times \mathbf{u}$). For instance, the resulting $(\bs{\omega} \times \mathbf{u}) \cdot \nabla \phi_y$ variable is often called as lift element. At $Re_c = 600$, the volume elements tend to dominate the contribution to the total force exerted over the wing.
		
	\section*{Acknowledgments}
	\label{sec:acknowledgments}
	We acknowledge the support from the US Air Force Office of Scientific Research (program manager: Dr. G. Abate, grant: FA9550-21-1-0174). We thank Profs. Yiyang Sun, Vassilios Theofilis, and Michael Amitay for enlightening discussions. We also thank the anonymous referees for providing constructive comments that significantly improved the quality of this manuscript. Computational resources were provided by the High Performance Computing Modernization Program at the US Department of Defense and the Texas Advanced Computing Center. 
	
	\section*{Declaration of interest}
	\label{sec:doi}
	The authors report no conflict of interest.
	
	\bibliography{taira_refs}

\begin{thebibliography}{99}
\expandafter\ifx\csname natexlab\endcsname\relax\def\natexlab#1{#1}\fi
\def\au#1{#1} \def\ed#1{#1} \def\yr#1{#1}\def\at#1{#1}\def\jt#1{\textit{#1}}
  \def\bt#1{#1}\def\bvol#1{\textbf{#1}} \def\vol#1{#1} \def\pg#1{#1}
  \def\publ#1{#1}\def\arxiv#1{#1}\def\org#1{#1}\def\st#1{\textit{#1}}

\bibitem[Amestoy {\em et~al.\/}(2001)Amestoy, Duff, L'Excellent \&
  Koster]{Amestoy:SIAM01}
{\sc \au{Amestoy, P.R.}, \au{Duff, I.~S.}, \au{L'Excellent, J.-Y.} \&
  \au{Koster, J.}} \yr{2001}  \at{A fully asynchronous multifrontal solver
  using distributed dynamic scheduling}.  \jt{SIAM J. Mat. Anal. App.}
  \bvol{23}~(1),  \pg{15--41}.

\bibitem[Amitay \& Glezer(2002)]{Amitay:AIAAJ02}
{\sc \au{Amitay, M.} \& \au{Glezer, A.}} \yr{2002}  \at{Role of actuation
  frequency in controlled flow reattachment over a stalled airfoil}.  \jt{AIAA
  J.}  \bvol{40}~(2),  \pg{209--216}.

\bibitem[Ananda {\em et~al.\/}(2015)Ananda, Sukumar \& Selig]{Ananda:AST15}
{\sc \au{Ananda, G.~K.}, \au{Sukumar, P.~P.} \& \au{Selig, M.~S.}} \yr{2015}
  \at{Measured aerodynamic characteristics of wings at low {R}eynolds numbers}.
   \jt{Aerosp. Sci. Technol.}  \bvol{42},  \pg{392--406}.

\bibitem[Anderson(2010)]{Anderson:10}
{\sc \au{Anderson, J.~D.}} \yr{2010} {\em Fundamentals of aerodynamics\/}.
  \publ{McGraw-Hill}.

\bibitem[Brandt {\em et~al.\/}(2023)Brandt, Mc{F}adden \&
  Bons]{Brandt:AIAA23control}
{\sc \au{Brandt, P.~J.}, \au{Mc{F}adden, E.~J.} \& \au{Bons, J.~P.}} \yr{2023}
  \at{Flow field characterization for swept wing active separation control}.
  \jt{AIAA Paper 2023--3285} .

\bibitem[Burtsev {\em et~al.\/}(2022)Burtsev, He, Hayostek, Zhang, Theofilis,
  Taira \& Amitay]{Burtsev:JFM22}
{\sc \au{Burtsev, A.}, \au{He, W.}, \au{Hayostek, S.}, \au{Zhang, K.},
  \au{Theofilis, V.}, \au{Taira, K.} \& \au{Amitay, M.}} \yr{2022}  \at{Linear
  modal instabilities around post-stall swept finite wings at low {R}eynolds
  numbers}.  \jt{J. Fluid Mech.}  \bvol{944},  \pg{A6}.

\bibitem[Chang(1992)]{Chang:PRSA92}
{\sc \au{Chang, C.-C.}} \yr{1992}  \at{Potential flow and forces for
  incompressible viscous flow}.  \jt{Proc. R. Soc. Lond. A}  \bvol{437}~(1901),
   \pg{517--525}.

\bibitem[Chen {\em et~al.\/}(2016)Chen, Bai \& Wang]{Chen:EJM2016singletailed}
{\sc \au{Chen, P.-W.}, \au{Bai, C.-J.} \& \au{Wang, W.-C.}} \yr{2016}
  \at{Experimental and numerical studies of low aspect ratio wing at critical
  {R}eynolds number}.  \jt{Eur. J. Mech. (B/Fluids)}  \bvol{59},
  \pg{161--168}.

\bibitem[Choi {\em et~al.\/}(2008)Choi, Jeon \& Kim]{Choi:ARFM08}
{\sc \au{Choi, H.}, \au{Jeon, W.-P.} \& \au{Kim, J.}} \yr{2008}  \at{Control of
  flow over a bluff body}.  \jt{Annu. Rev. Fluid Mech.}  \bvol{40},
  \pg{113--139}.

\bibitem[Chu(1965)]{Chu:Acta65}
{\sc \au{Chu, B.-T.}} \yr{1965}  \at{On the energy transfer to small
  disturbances in fluid flow (part {I})}.  \jt{Acta Mechanica}  \bvol{1}~(3),
  \pg{215--234}.

\bibitem[Dallmann(1988)]{DallmannFDR1988}
{\sc \au{Dallmann, U.~C.}} \yr{1988}  \at{Three-dimensional vortex structures
  and vorticity topology}.  \jt{Fluid Dyn. Res.}  \bvol{3}~(1-4),
  \pg{183--189}.

\bibitem[D{\'e}lery(2001)]{Delery:AR01}
{\sc \au{D{\'e}lery, J.~M.}} \yr{2001}  \at{Robert {L}egendre and {H}enri
  {W}erl{\'e}: toward the elucidation of three-dimensional separation}.
  \jt{Annu. Rev. Fluid Mech.}  \bvol{33},  \pg{128}.

\bibitem[Devenport {\em et~al.\/}(1996)Devenport, Rife, Liapis \&
  Follin]{Devenport:JFM96}
{\sc \au{Devenport, W.~J.}, \au{Rife, M.~C.}, \au{Liapis, S.~I.} \& \au{Follin,
  G.~J.}} \yr{1996}  \at{The structure and development of a wing-tip vortex}.
  \jt{J. Fluid Mech.}  \bvol{312},  \pg{67--106}.

\bibitem[Dong {\em et~al.\/}(2020)Dong, Choi \& Mao]{Dong:EF20}
{\sc \au{Dong, L.}, \au{Choi, K.-S.} \& \au{Mao, X.}} \yr{2020}  \at{Interplay
  of the leading-edge vortex and the tip vortex of a low-aspect-ratio thin
  wing}.  \jt{Exp. Fluids}  \bvol{61}~(9),  \pg{1--15}.

\bibitem[Edstrand {\em et~al.\/}(2018{\natexlab{{\em a\/}}})Edstrand, Schmid,
  Taira \& {Cattafesta III}]{Edstrand:JFM18a}
{\sc \au{Edstrand, A.~M.}, \au{Schmid, P.~J.}, \au{Taira, K.} \&
  \au{{Cattafesta III}, L.~N.}} \yr{2018{\natexlab{{\em a\/}}}}  \at{A parallel
  stability analysis of a trailing vortex wake}.  \jt{J. Fluid Mech.}
  \bvol{837},  \pg{858}.

\bibitem[Edstrand {\em et~al.\/}(2018{\natexlab{{\em b\/}}})Edstrand, Sun,
  Schmid, Taira \& Cattafesta]{Edstrand:JFM18b}
{\sc \au{Edstrand, A.~M.}, \au{Sun, Y.}, \au{Schmid, P.~J.}, \au{Taira, K.} \&
  \au{Cattafesta, L.~N.}} \yr{2018{\natexlab{{\em b\/}}}}  \at{Active
  attenuation of a trailing vortex inspired by a parabolized stability
  analysis}.  \jt{J. Fluid Mech.}  \bvol{855},  \pg{R2}.

\bibitem[Eisenlohr \& Eckelmann(1989)]{Eisenlohr:PF89vortexDislocation}
{\sc \au{Eisenlohr, H.} \& \au{Eckelmann, H.}} \yr{1989}  \at{Vortex splitting
  and its consequences in the vortex street wake of cylinders at low {R}eynolds
  number}.  \jt{Phys. Fluids A}  \bvol{1}~(2),  \pg{189--192}.

\bibitem[Eldredge \& Jones(2019)]{Eldredge:ARFM19}
{\sc \au{Eldredge, J.~D.} \& \au{Jones, A.~R.}} \yr{2019}  \at{Leading-edge
  vortices: mechanics and modeling}.  \jt{Annu. Rev. Fluid Mech.}  \bvol{51},
  \pg{75--104}.

\bibitem[Ellington {\em et~al.\/}(1996)Ellington, Berg, Willmott \&
  Thomas]{Ellington:Nature96}
{\sc \au{Ellington, C.~P.}, \au{Berg, C. Van~Den}, \au{Willmott, A.~P.} \&
  \au{Thomas, A. L.~R.}} \yr{1996}  \at{Leading-edge vortices in insect
  flight}.  \jt{Nature}  \bvol{384}~(6610),  \pg{626--630}.

\bibitem[Fage \& Johansen(1927)]{FageJohanssen:PRSA27}
{\sc \au{Fage, A.} \& \au{Johansen, F.~C.}} \yr{1927}  \at{On the flow of air
  behind an inclined flat plate of infinite span}.  \jt{Proc. R. Soc. Lond. A}
  \bvol{116}~(773),  \pg{170--197}.

\bibitem[{Fosas de Pando} \& Schmid(2017)]{Fosas:JoT17}
{\sc \au{{Fosas de Pando}, M.} \& \au{Schmid, P.~J.}} \yr{2017}  \at{Optimal
  frequency-response sensitivity of compressible flow over roughness elements}.
   \jt{J. Turb.}  \bvol{18}~(4),  \pg{338--351}.

\bibitem[Francis \& Kennedy(1979)]{Francis:JA79}
{\sc \au{Francis, M.~S.} \& \au{Kennedy, D.~A.}} \yr{1979}  \at{Formation of a
  trailing vortex}.  \jt{J. Aircraft}  \bvol{16}~(3),  \pg{148--154}.

\bibitem[Freund(1997)]{Freund:AIAAJ97}
{\sc \au{Freund, J.~B.}} \yr{1997}  \at{Proposed inflow/outflow boundary
  condition for direct computation of aerodynamic sound}.  \jt{AIAA J.}
  \bvol{35}~(4),  \pg{740--742}.

\bibitem[Freymuth {\em et~al.\/}(1987)Freymuth, Finaish \&
  Bank]{Freymuth:AIAAJ87}
{\sc \au{Freymuth, P.}, \au{Finaish, F.} \& \au{Bank, W.}} \yr{1987}
  \at{Further visualization of combined wing tip and starting vortex systems}.
  \jt{AIAA J.}  \bvol{25}~(9),  \pg{1153--1159}.

\bibitem[{Gopalakrishnan Meena} {\em et~al.\/}(2017){Gopalakrishnan Meena},
  Taira \& Asai]{Gopalakrishnan:AIAAJ17}
{\sc \au{{Gopalakrishnan Meena}, M.}, \au{Taira, K.} \& \au{Asai, K.}}
  \yr{2017}  \at{Airfoil-wake modification with {G}urney flap at low {R}eynolds
  number}.  \jt{AIAA J.}  \bvol{56}~(4),  \pg{1348--1359}.

\bibitem[Green \& Acosta(1991)]{Green:JFM91}
{\sc \au{Green, S.~I.} \& \au{Acosta, A.~J.}} \yr{1991}  \at{Unsteady flow in
  trailing vortices}.  \jt{J. Fluid Mech.}  \bvol{227},  \pg{107--134}.

\bibitem[Greenblatt(2012)]{Greenblatt:AIAAJ12}
{\sc \au{Greenblatt, D.}} \yr{2012}  \at{Fluidic control of a wing tip vortex}.
   \jt{AIAA J.}  \bvol{50}~(2),  \pg{375--386}.

\bibitem[Greenblatt \& Wygnanski(2000)]{Greenblatt:PAS00}
{\sc \au{Greenblatt, D.} \& \au{Wygnanski, I.~J.}} \yr{2000}  \at{The control
  of flow separation by periodic excitation}.  \jt{Prog. Aero. Sci.}
  \bvol{36},  \pg{487--545}.

\bibitem[Gross {\em et~al.\/}(2024)Gross, Marks \&
  Sondergaard]{Gross:AIAAJ2024resolventcontrol}
{\sc \au{Gross, A.}, \au{Marks, C.} \& \au{Sondergaard, R.}} \yr{2024}
  \at{Laminar separation control for {E}ppler 387 airfoil based on resolvent
  analysis}.  \jt{AIAA J.}  \pg{pp. 1--16}.

\bibitem[Gursul {\em et~al.\/}(2007)Gursul, Vardaki, Margaris \&
  Wang]{Gursul:AFC07tipcontrol}
{\sc \au{Gursul, I.}, \au{Vardaki, E.}, \au{Margaris, P.} \& \au{Wang, Z.}}
  \yr{2007} Control of wing vortices.  \bt{In {\em Active Flow Control\/}},
  \pg{pp. 137--151}. Springer.

\bibitem[Gursul \& Wang(2018)]{Gursul:AIAAJ18}
{\sc \au{Gursul, I.} \& \au{Wang, Z.}} \yr{2018}  \at{Flow control of tip/edge
  vortices}.  \jt{AIAA J.}  \bvol{56}~(5),  \pg{1731--1749}.

\bibitem[Ham \& Iaccarino(2004)]{Ham:CTR04energy}
{\sc \au{Ham, F.} \& \au{Iaccarino, G.}} \yr{2004}  \at{Energy conservation in
  collocated discretization schemes on unstructured meshes}.  \jt{Annual
  Research Briefs, Center for Turbulence Research}  \pg{pp. 3--14}.

\bibitem[Ham {\em et~al.\/}(2006)Ham, Mattsson \& Iaccarino]{Ham:CTR06accurate}
{\sc \au{Ham, F.}, \au{Mattsson, K.} \& \au{Iaccarino, G.}} \yr{2006}
  \at{Accurate and stable finite volume operators for unstructured flow
  solvers}.  \jt{Annual Research Briefs, Center for Turbulence Research}
  \pg{pp. 243--261}.

\bibitem[Hayostek {\em et~al.\/}(2022)Hayostek, Ottinger, Neal \&
  Amitay]{Hayostek:AIAA22}
{\sc \au{Hayostek, S.}, \au{Ottinger, J.}, \au{Neal, J.} \& \au{Amitay, M.}}
  \yr{2022}  \at{Experimental investigation on the effect of sweep and taper on
  low {R}eynolds number finite wings}.  \jt{AIAA Paper 2022--0467} .

\bibitem[He {\em et~al.\/}(2017)He, Gioria, P{\'e}rez \&
  V.~Theofilis]{He:JFM17}
{\sc \au{He, W.}, \au{Gioria, R.~S.}, \au{P{\'e}rez, J.~M.} \&
  \au{V.~Theofilis, Vassilis}} \yr{2017}  \at{Linear instability of low
  {R}eynolds number massively separated flow around three {NACA} airfoils}.
  \jt{J. Fluid Mech.}  \bvol{811},  \pg{701--741}.

\bibitem[Houtman {\em et~al.\/}(2023)Houtman, Timme \&
  Sharma]{Houtman:Flow23resolvent}
{\sc \au{Houtman, J.}, \au{Timme, S.} \& \au{Sharma, A.}} \yr{2023}
  \at{Resolvent analysis of a finite wing in transonic flow}.  \jt{Flow}
  \bvol{3},  \pg{E14}.

\bibitem[Hunt {\em et~al.\/}(1978)Hunt, Abel, Peterka \& Woo]{HuntEtAl:JFM1978}
{\sc \au{Hunt, J. C.~R.}, \au{Abel, C.~J.}, \au{Peterka, J.~A.} \& \au{Woo,
  H.}} \yr{1978}  \at{Kinematic studies of the flows around free or
  surface-mounted obstacles; applying topology to flow visualization}.  \jt{J.
  Fluid Mech.}  \bvol{86}~(1),  \pg{179--200}.

\bibitem[Hunt {\em et~al.\/}(1988)Hunt, Wray \& Moin]{Hunt:CTR1988eddies}
{\sc \au{Hunt, J. C.~R.}, \au{Wray, A.~A.} \& \au{Moin, P.}} \yr{1988}
  \at{Eddies, streams, and convergence zones in turbulent flows}.  \jt{Studying
  turbulence using numerical simulation databases, 2. Proceedings of the 1988
  summer program} .

\bibitem[Jeong \& Hussain(1995)]{Jeong:JFM95}
{\sc \au{Jeong, J.} \& \au{Hussain, F.}} \yr{1995}  \at{On the identification
  of a vortex}.  \jt{J. Fluid Mech.}  \bvol{285},  \pg{69--94}.

\bibitem[Jin {\em et~al.\/}(2020)Jin, Illingworth \&
  Sandberg]{Jin:JFM2020feedbackresolvent}
{\sc \au{Jin, B.}, \au{Illingworth, S.~J.} \& \au{Sandberg, R.~D.}} \yr{2020}
  \at{Feedback control of vortex shedding using a resolvent-based modelling
  approach}.  \jt{J. Fluid Mech.}  \bvol{897},  \pg{A26}.

\bibitem[Jovanovi{\'c} \& Bamieh(2005)]{Jovanovic:JFM05}
{\sc \au{Jovanovi{\'c}, M.~R.} \& \au{Bamieh, B.}} \yr{2005}  \at{Componentwise
  energy amplification in channel flows}.  \jt{J. Fluid Mech.}  \bvol{534},
  \pg{145--183}.

\bibitem[Katz \& {Bueno Galdo}(1989)]{Katz:JA1989roughness}
{\sc \au{Katz, J.} \& \au{{Bueno Galdo}, J.}} \yr{1989}  \at{Effect of
  roughness on rollup of tip vortices on a rectangular hydrofoil}.  \jt{J.
  Aircraft}  \bvol{26}~(3),  \pg{247--253}.

\bibitem[Lee {\em et~al.\/}(2012)Lee, Hsieh, Chang \& Chu]{Lee:JFM12}
{\sc \au{Lee, J.-J.}, \au{Hsieh, C.-T.}, \au{Chang, C.-C.} \& \au{Chu, C.-C.}}
  \yr{2012}  \at{Vorticity forces on an impulsively started finite plate}.
  \jt{J. Fluid Mech.}  \bvol{694},  \pg{464--492}.

\bibitem[Lin {\em et~al.\/}(2023)Lin, Tsai \& Tsai]{Lin:JFM2023flow}
{\sc \au{Lin, C.-T.}, \au{Tsai, M.-L.} \& \au{Tsai, H.-C.}} \yr{2023}  \at{Flow
  control of a plunging cylinder based on resolvent analysis}.  \jt{J. Fluid
  Mech.}  \bvol{967},  \pg{A41}.

\bibitem[Liu {\em et~al.\/}(2021)Liu, Sun, Yeh, Ukeiley, Cattafesta \&
  Taira]{Liu:JFM21}
{\sc \au{Liu, Q.}, \au{Sun, Y.}, \au{Yeh, C.-A.}, \au{Ukeiley, L.~S.},
  \au{Cattafesta, L.~N.} \& \au{Taira, K.}} \yr{2021}  \at{Unsteady control of
  supersonic turbulent cavity flow based on resolvent analysis}.  \jt{J. Fluid
  Mech.}  \bvol{925},  \pg{A5}.

\bibitem[Luhar {\em et~al.\/}(2014)Luhar, Sharma \&
  McKeon]{Luhar:JFM14opposition}
{\sc \au{Luhar, M.}, \au{Sharma, A.~S.} \& \au{McKeon, B.~J.}} \yr{2014}
  \at{Opposition control within the resolvent analysis framework}.  \jt{J.
  Fluid Mech.}  \bvol{749},  \pg{597--626}.

\bibitem[Mayer \& Powell(1992)]{Mayer:JFM92viscous}
{\sc \au{Mayer, E.~W.} \& \au{Powell, K.~G.}} \yr{1992}  \at{Viscous and
  inviscid instabilities of a trailing vortex}.  \jt{J. Fluid Mech.}
  \bvol{245},  \pg{91--114}.

\bibitem[Mc{F}adden {\em et~al.\/}(2022)Mc{F}adden, Brandt \&
  Bons]{Mcfadden:AIAA22control}
{\sc \au{Mc{F}adden, E.~J.}, \au{Brandt, P.~J.} \& \au{Bons, J.~P.}} \yr{2022}
  \at{Swept wing active flow control using a streamwise row of vortex
  generating jets}.  \jt{AIAA Paper 2022--1546} .

\bibitem[Mc{K}eon \& Sharma(2010)]{McKeon:JFM10}
{\sc \au{Mc{K}eon, B.~J.} \& \au{Sharma, A.~S.}} \yr{2010}  \at{A
  critical-layer framework for turbulent pipe flow}.  \jt{J. Fluid Mech.}
  \bvol{658},  \pg{336--382}.

\bibitem[Menon \& Mittal(2021)]{Menon:JCP21}
{\sc \au{Menon, K.} \& \au{Mittal, R.}} \yr{2021}  \at{Quantitative analysis of
  the kinematics and induced aerodynamic loading of individual vortices in
  vortex-dominated flows: a computation and data-driven approach}.  \jt{J.
  Comput. Phys.}  \bvol{443},  \pg{110515}.

\bibitem[Mizoguchi {\em et~al.\/}(2016)Mizoguchi, Kajikawa \&
  Itoh]{MizoguchiTJSA:2016aerodynamic}
{\sc \au{Mizoguchi, M.}, \au{Kajikawa, Y.} \& \au{Itoh, H.}} \yr{2016}
  \at{Aerodynamic characteristics of low-aspect-ratio wings with various aspect
  ratios in low {R}eynolds number flows}.  \jt{Trans. Japan Soc. Aero. S Sci.}
  \bvol{59}~(2),  \pg{56--63}.

\bibitem[Moarref {\em et~al.\/}(2013)Moarref, Sharma, Tropp \&
  McKeon]{Moarref:JFM13}
{\sc \au{Moarref, R.}, \au{Sharma, A.~S.}, \au{Tropp, J.~A.} \& \au{McKeon,
  B.~J.}} \yr{2013}  \at{Model-based scaling of the streamwise energy density
  in high-{R}eynolds-number turbulent channels}.  \jt{J. Fluid Mech.}
  \bvol{734},  \pg{275--316}.

\bibitem[Mullenix {\em et~al.\/}(2013)Mullenix, Gaitonde \&
  Visbal]{Mullenix:AIAAJ2013bodyforce}
{\sc \au{Mullenix, N.~J.}, \au{Gaitonde, D.~V.} \& \au{Visbal, M.~R.}}
  \yr{2013}  \at{Spatially developing supersonic turbulent boundary layer with
  a body-force-based method}.  \jt{AIAA J.}  \bvol{51}~(8),  \pg{1805--1819}.

\bibitem[Nastro {\em et~al.\/}(2023)Nastro, Robinet, Loiseau, Passaggia \&
  Mazellier]{Nastro:JFM23}
{\sc \au{Nastro, G.}, \au{Robinet, J.-C.}, \au{Loiseau, J.-C.}, \au{Passaggia,
  P.-Y.} \& \au{Mazellier, N.}} \yr{2023}  \at{Global stability, sensitivity
  and passive control of low-{R}eynolds-number flows around {NACA} 4412 swept
  wings}.  \jt{J. Fluid Mech.}  \bvol{957},  \pg{A5}.

\bibitem[Navrose {\em et~al.\/}(2019)Navrose, Brion \& Jacquin]{Navrose:JFM19}
{\sc \au{Navrose}, \au{Brion, V.} \& \au{Jacquin, L.}} \yr{2019}  \at{Transient
  growth in the near wake region of the flow past a finite span wing}.  \jt{J.
  Fluid Mech.}  \bvol{866},  \pg{399--430}.

\bibitem[Neal \& Amitay(2023)]{Neal:PRF23}
{\sc \au{Neal, J.~M.} \& \au{Amitay, M.}} \yr{2023}  \at{Three-dimensional
  separation over unswept cantilevered wings at a moderate {R}eynolds number}.
  \jt{Phys. Rev. Fluids}  \bvol{8},  \pg{014703}.

\bibitem[Okamoto \& Azuma(2011)]{Okamoto:AIAAJ2011aerodynamic}
{\sc \au{Okamoto, M.} \& \au{Azuma, A.}} \yr{2011}  \at{Aerodynamic
  characteristics at low {R}eynolds number for wings of various planforms}.
  \jt{AIAA J.}  \bvol{49}~(6),  \pg{1135--1150}.

\bibitem[Okamoto {\em et~al.\/}(2019)Okamoto, Sasaki, Kamikubo \&
  Fujii]{Okamoto:TJSA2019disappearance}
{\sc \au{Okamoto, M.}, \au{Sasaki, D.}, \au{Kamikubo, M.} \& \au{Fujii, R.}}
  \yr{2019}  \at{Disappearance of vortex lift in low-aspect-ratio wings at
  very-low {R}eynolds numbers}.  \jt{Trans. Japan Soc. Aero. S Sci.}
  \bvol{62}~(6),  \pg{310--317}.

\bibitem[Pandi \& Mittal(2023)]{Pandi:JFM23}
{\sc \au{Pandi, J. S.~S.} \& \au{Mittal, S.}} \yr{2023}  \at{Streamwise
  vortices, cellular shedding and force coefficients on finite wing at low
  {R}eynolds number}.  \jt{J. Fluid Mech.}  \bvol{958},  \pg{A10}.

\bibitem[Pauley {\em et~al.\/}(1990)Pauley, Moin \& {R}eynolds]{Pauley:JFM90}
{\sc \au{Pauley, L.~L.}, \au{Moin, P.} \& \au{{R}eynolds, W.~C.}} \yr{1990}
  \at{The structure of two-dimensional separation}.  \jt{J. Fluid Mech.}
  \bvol{220},  \pg{397--411}.

\bibitem[Prasad \& Unnikrishnan(2023)]{Prasad:JFM2023plasma}
{\sc \au{Prasad, A. L.~N.} \& \au{Unnikrishnan, S.}} \yr{2023}  \at{Effect of
  plasma actuator-based control on flow-field and acoustics of supersonic
  rectangular jets}.  \jt{J. Fluid Mech.}  \bvol{964},  \pg{A11}.

\bibitem[Quartapelle \& Napolitano(1983)]{Quartapelle:AIAAJ83}
{\sc \au{Quartapelle, L.} \& \au{Napolitano, M.}} \yr{1983}  \at{Force and
  moment in incompressible flows}.  \jt{AIAA J.}  \bvol{21}~(6),
  \pg{911--913}.

\bibitem[Ribeiro {\em et~al.\/}(2020{\natexlab{{\em a\/}}})Ribeiro, Frank \&
  Franck]{RibeiroBernardo:JFS20reynoldseffects}
{\sc \au{Ribeiro, B. L.~R.}, \au{Frank, S.~L.} \& \au{Franck, J.~A.}}
  \yr{2020{\natexlab{{\em a\/}}}}  \at{Vortex dynamics and {R}eynolds number
  effects of an oscillating hydrofoil in energy harvesting mode}.  \jt{J.
  Fluids Struct.}  \bvol{94},  \pg{102888}.

\bibitem[Ribeiro {\em et~al.\/}(2023{\natexlab{{\em a\/}}})Ribeiro, Neal,
  Burtsev, Amitay, Theofilis \& Taira]{Ribeiro:JFM2023tapered}
{\sc \au{Ribeiro, J. H.~M.}, \au{Neal, J.}, \au{Burtsev, A.}, \au{Amitay, M.},
  \au{Theofilis, V.} \& \au{Taira, K.}} \yr{2023{\natexlab{{\em a\/}}}}
  \at{Laminar post-stall wakes of tapered swept wings}.  \jt{J. Fluid Mech.}
  \bvol{976},  \pg{A6}.

\bibitem[Ribeiro \& Taira(2023)]{Ribeiro:AIAA23resolvent}
{\sc \au{Ribeiro, J. H.~M.} \& \au{Taira, K.}} \yr{2023}  \at{Resolvent-based
  analysis of low-{R}eynolds-number separated flows around tapered wings}.
  \jt{AIAA Paper 2023--4354} .

\bibitem[Ribeiro {\em et~al.\/}(2020{\natexlab{{\em b\/}}})Ribeiro, Yeh \&
  Taira]{Ribeiro:PRF20}
{\sc \au{Ribeiro, J. H.~M.}, \au{Yeh, C.-A.} \& \au{Taira, K.}}
  \yr{2020{\natexlab{{\em b\/}}}}  \at{Randomized resolvent analysis}.
  \jt{Phys. Rev. Fluids}  \bvol{5}~(3),  \pg{033902}.

\bibitem[Ribeiro {\em et~al.\/}(2023{\natexlab{{\em b\/}}})Ribeiro, Yeh \&
  Taira]{Ribeiro:JFM23triglobal}
{\sc \au{Ribeiro, J. H.~M.}, \au{Yeh, C.-A.} \& \au{Taira, K.}}
  \yr{2023{\natexlab{{\em b\/}}}}  \at{Triglobal resolvent analysis of
  swept-wing wakes}.  \jt{J. Fluid Mech.}  \bvol{954},  \pg{A42}.

\bibitem[Ribeiro {\em et~al.\/}(2022)Ribeiro, Yeh, Zhang \&
  Taira]{Ribeiro:JFM22}
{\sc \au{Ribeiro, J. H.~M.}, \au{Yeh, C.-A.}, \au{Zhang, K.} \& \au{Taira, K.}}
  \yr{2022}  \at{Wing sweep effects on laminar separated flows}.  \jt{J. Fluid
  Mech.}  \bvol{950},  \pg{A23}.

\bibitem[Ricciardi {\em et~al.\/}(2022)Ricciardi, Wolf \&
  Taira]{Ricciardi:JFM22}
{\sc \au{Ricciardi, T.~R.}, \au{Wolf, W.~R.} \& \au{Taira, K.}} \yr{2022}
  \at{Transition, intermittency and phase interference effects in airfoil
  secondary tones and acoustic feedback loop}.  \jt{J. Fluid Mech.}
  \bvol{937}.

\bibitem[Schmid \& Brandt(2014)]{Schmid:AMR14}
{\sc \au{Schmid, P.~J.} \& \au{Brandt, L.}} \yr{2014}  \at{Analysis of fluid
  systems: Stability, receptivity, sensitivity}.  \jt{Applied Mechanics
  Reviews}  \bvol{66}~(2), 024803.

\bibitem[Schmidt {\em et~al.\/}(2018)Schmidt, Towne, Rigas, Colonius \&
  Br{\`e}s]{Schmidt:JFM18}
{\sc \au{Schmidt, O.~T.}, \au{Towne, A.}, \au{Rigas, G.}, \au{Colonius, T.} \&
  \au{Br{\`e}s, G.~A.}} \yr{2018}  \at{Spectral analysis of jet turbulence}.
  \jt{J. Fluid Mech.}  \bvol{855},  \pg{953--982}.

\bibitem[Seifert {\em et~al.\/}(1996)Seifert, Darabi \&
  Wygnanski]{Seifert:JA96}
{\sc \au{Seifert, A.}, \au{Darabi, A.} \& \au{Wygnanski, I.}} \yr{1996}
  \at{Delay of airfoil stall by periodic excitation}.  \jt{J. Aircraft}
  \bvol{33}~(4),  \pg{691--698}.

\bibitem[Shyy {\em et~al.\/}(2002)Shyy, Jayaraman \&
  Andersson]{Shyy:JAP2002bodyforce}
{\sc \au{Shyy, W.}, \au{Jayaraman, B.} \& \au{Andersson, A.}} \yr{2002}
  \at{Modeling of glow discharge-induced fluid dynamics}.  \jt{J. Appl. Phys.}
  \bvol{92}~(11),  \pg{6434--6443}.

\bibitem[Shyy {\em et~al.\/}(2008)Shyy, Lian, Tang, Viieru \& Liu]{Shyy08}
{\sc \au{Shyy, W.}, \au{Lian, Y.}, \au{Tang, J.}, \au{Viieru, D.} \& \au{Liu,
  H.}} \yr{2008} {\em Aerodynamics of low {R}eynolds number flyers\/}.
  \publ{Cambridge Univ. Press}.

\bibitem[Skene {\em et~al.\/}(2022)Skene, Ribeiro \& Taira]{SkeneRibeiro:tools}
{\sc \au{Skene, C.~S.}, \au{Ribeiro, J. H.~M.} \& \au{Taira, K.}} \yr{2022}
  csskene/linear-analysis-tools: Initial release.
  https://doi.org/10.5281/zenodo.6550726.

\bibitem[Spalart(1998)]{Spalart:ARFM98}
{\sc \au{Spalart, P.~S.}} \yr{1998}  \at{Airplane trailing vortices}.
  \jt{Annu. Rev. Fluid Mech.}  \bvol{30},  \pg{107}.

\bibitem[Taira {\em et~al.\/}(2017)Taira, Brunton, Dawson, Rowley, Colonius,
  McKeon, Schmidt, Gordeyev, Theofilis \& Ukeiley]{Taira:AIAAJ17}
{\sc \au{Taira, K.}, \au{Brunton, S.~L.}, \au{Dawson, S. T.~M.}, \au{Rowley,
  C.~W.}, \au{Colonius, T.}, \au{McKeon, B.~J.}, \au{Schmidt, O.~T.},
  \au{Gordeyev, S.}, \au{Theofilis, V.} \& \au{Ukeiley, L.~S.}} \yr{2017}
  \at{Modal analysis of fluid flows: An overview}.  \jt{AIAA J.}
  \bvol{55}~(12),  \pg{4013--4041}.

\bibitem[Taira \& Colonius(2009{\natexlab{{\em a\/}}})]{Taira:AIAAJ09}
{\sc \au{Taira, K.} \& \au{Colonius, T.}} \yr{2009{\natexlab{{\em a\/}}}}
  \at{Effect of tip vortices and low-{R}eynolds-number poststall flow control}.
   \jt{AIAA J.}  \bvol{47}~(3),  \pg{749--756}.

\bibitem[Taira \& Colonius(2009{\natexlab{{\em b\/}}})]{Taira:JFM09}
{\sc \au{Taira, K.} \& \au{Colonius, T.}} \yr{2009{\natexlab{{\em b\/}}}}
  \at{Three-dimensional flows around low-aspect-ratio flat-plate wings at low
  {R}eynolds numbers}.  \jt{J. Fluid Mech.}  \bvol{623},  \pg{187--207}.

\bibitem[Toppings \& Yarusevych(2021)]{Toppings:JFM21}
{\sc \au{Toppings, C.~E.} \& \au{Yarusevych, S.}} \yr{2021}  \at{Structure and
  dynamics of a laminar separation bubble near a wingtip}.  \jt{J. Fluid Mech.}
   \bvol{929},  \pg{A39}.

\bibitem[Toppings \& Yarusevych(2022)]{Toppings:JFM22}
{\sc \au{Toppings, C.~E.} \& \au{Yarusevych, S.}} \yr{2022}  \at{Structure and
  dynamics of a laminar separation bubble near a wing root: towards
  reconstructing the complete {LSB} topology on a finite wing}.  \jt{J. Fluid
  Mech.}  \bvol{944},  \pg{A14}.

\bibitem[Torres \& Mueller(2004)]{Torres:AIAAJ04}
{\sc \au{Torres, G.~E.} \& \au{Mueller, T.~J.}} \yr{2004}  \at{Low-aspect-ratio
  aerodynamics at low {R}eynolds numbers}.  \jt{AIAA J.}  \bvol{42}~(5),
  \pg{865--873}.

\bibitem[Trefethen {\em et~al.\/}(1993)Trefethen, Trefethen, Reddy \&
  Driscoll]{Trefethen:93}
{\sc \au{Trefethen, L.~N.}, \au{Trefethen, A.~E.}, \au{Reddy, S.~C.} \&
  \au{Driscoll, T.~A.}} \yr{1993}  \at{Hydrodynamic stability without
  eigenvalues}.  \jt{Science}  \bvol{261}~(5121),  \pg{578--584}.

\bibitem[Unnikrishnan(2023)]{Unnikrishnan:PAS23recent}
{\sc \au{Unnikrishnan, S.}} \yr{2023}  \at{Recent advances in feature
  extraction techniques for high-speed flowfields}.  \jt{Prog. Aero. Sci.}
  \pg{p. 100918}.

\bibitem[Videler {\em et~al.\/}(2004)Videler, Stamhuis \&
  Povel]{Videler:Science04}
{\sc \au{Videler, J.~J.}, \au{Stamhuis, E.~J.} \& \au{Povel, G. D.~E.}}
  \yr{2004}  \at{Leading-edge vortex lifts swifts}.  \jt{Science}
  \bvol{306}~(5703),  \pg{1960--1962}.

\bibitem[Viieru {\em et~al.\/}(2005)Viieru, Albertani, Shyy \&
  Ifju]{ViieruShyy:JA05}
{\sc \au{Viieru, D.}, \au{Albertani, R.}, \au{Shyy, W.} \& \au{Ifju, P.~G.}}
  \yr{2005}  \at{Effect of tip vortex on wing aerodynamics of micro air
  vehicles}.  \jt{J. Aircraft}  \bvol{42}~(6),  \pg{1530--1536}.

\bibitem[Visbal {\em et~al.\/}(2013)Visbal, Yilmaz \&
  Rockwell]{Visbal:2013yilmazrockwell}
{\sc \au{Visbal, M.}, \au{Yilmaz, T.~O.} \& \au{Rockwell, D.}} \yr{2013}
  \at{Three-dimensional vortex formation on a heaving low-aspect-ratio wing:
  Computations and experiments}.  \jt{J. Fluids Struct.}  \bvol{38},
  \pg{58--76}.

\bibitem[Visbal(2012)]{Visbal:AIAA12}
{\sc \au{Visbal, M.~R.}} \yr{2012}  \bt{Flow structure and unsteady loading
  over a pitching low-aspect-ratio wing}. {AIAA 2012-3279}.

\bibitem[Waindim \& Gaitonde(2016)]{WaindimGaitonde:JCP2016bodyforce}
{\sc \au{Waindim, M.} \& \au{Gaitonde, D.~V.}} \yr{2016}  \at{A body-force
  based method to generate supersonic equilibrium turbulent boundary layer
  profiles}.  \jt{J. Comput. Phys.}  \bvol{304},  \pg{1--26}.

\bibitem[Williamson(1989)]{Williamson:JFM89oblique}
{\sc \au{Williamson, C. H.~K.}} \yr{1989}  \at{Oblique and parallel modes of
  vortex shedding in the wake of a circular cylinder at low {R}eynolds
  numbers}.  \jt{J. Fluid Mech.}  \bvol{206},  \pg{579--627}.

\bibitem[Yarusevych {\em et~al.\/}(2009)Yarusevych, Sullivan \&
  Kawall]{Yarusevych:JFM09}
{\sc \au{Yarusevych, S.}, \au{Sullivan, P.~E.} \& \au{Kawall, J.~G.}} \yr{2009}
   \at{On vortex shedding from an airfoil in low-{R}eynolds-number flows}.
  \jt{J. Fluid Mech.}  \bvol{632},  \pg{245--271}.

\bibitem[Yeh {\em et~al.\/}(2017)Yeh, Munday \& Taira]{Yeh:JFM2017laminar}
{\sc \au{Yeh, C.-A.}, \au{Munday, P.~M.} \& \au{Taira, K.}} \yr{2017}
  \at{Laminar free shear layer modification using localized periodic heating}.
  \jt{J. Fluid Mech.}  \bvol{822},  \pg{561--589}.

\bibitem[Yeh \& Taira(2019)]{Yeh:JFM19}
{\sc \au{Yeh, C.-A.} \& \au{Taira, K.}} \yr{2019}  \at{Resolvent-analysis-based
  design of airfoil separation control}.  \jt{J. Fluid Mech.}  \bvol{867},
  \pg{572--610}.

\bibitem[Yilmaz \& Rockwell(2012)]{Yilmaz:JFM12}
{\sc \au{Yilmaz, T.~O.} \& \au{Rockwell, D.}} \yr{2012}  \at{Flow structure on
  finite-span wings due to pitch-up motion}.  \jt{J. Fluid Mech.}  \bvol{691},
  \pg{518--545}.

\bibitem[Zhang {\em et~al.\/}(2020{\natexlab{{\em a\/}}})Zhang, Hayostek,
  Amitay, Burstev, Theofilis \& Taira]{Zhang:JFM20b}
{\sc \au{Zhang, K.}, \au{Hayostek, S.}, \au{Amitay, M.}, \au{Burstev, A.},
  \au{Theofilis, V.} \& \au{Taira, K.}} \yr{2020{\natexlab{{\em a\/}}}}
  \at{Laminar separated flows over finite-aspect-ratio swept wings}.  \jt{J.
  Fluid Mech.}  \bvol{905},  \pg{R1}.

\bibitem[Zhang {\em et~al.\/}(2020{\natexlab{{\em b\/}}})Zhang, Hayostek,
  Amitay, He, Theofilis \& Taira]{Zhang:JFM20}
{\sc \au{Zhang, K.}, \au{Hayostek, S.}, \au{Amitay, M.}, \au{He, W.},
  \au{Theofilis, V.} \& \au{Taira, K.}} \yr{2020{\natexlab{{\em b\/}}}}  \at{On
  the formation of three-dimensional separated flows over wings under tip
  effects}.  \jt{J. Fluid Mech.}  \bvol{895},  \pg{A9}.

\bibitem[Zhang \& Taira(2022)]{Zhang:PRF22}
{\sc \au{Zhang, K.} \& \au{Taira, K.}} \yr{2022}  \at{Laminar vortex dynamics
  around forward-swept wings}.  \jt{Phys. Rev. Fluids}  \bvol{7}~(2),
  \pg{024704}.

\bibitem[Zhu {\em et~al.\/}(2024)Zhu, Wang \& Liu]{Zhu:JFM2024tip}
{\sc \au{Zhu, Y.}, \au{Wang, J.} \& \au{Liu, J.}} \yr{2024}  \at{Tip effects on
  three-dimensional flow structures over low-aspect-ratio plates: mechanisms of
  spanwise fluid transport}.  \jt{J. Fluid Mech.}  \bvol{983},  \pg{A35}.

\bibitem[Zhu {\em et~al.\/}(2023)Zhu, Wang, Xu, Qu \& Long]{Zhu:JFM2023swallow}
{\sc \au{Zhu, Y.}, \au{Wang, J.}, \au{Xu, Y.}, \au{Qu, Y.} \& \au{Long, Y.}}
  \yr{2023}  \at{Swallow-tailed separation bubble on a low-aspect-ratio
  trapezoidal plate: effects of near-wall spanwise flow}.  \jt{J. Fluid Mech.}
  \bvol{965},  \pg{A12}.

\end{thebibliography}
	\bibliographystyle{unsrt} 
	
\end{document}